\documentclass[12pt,fleqn,a4paper]{article} 
\usepackage{epsfig,graphics,graphicx}
\usepackage{clshan-math,clshan-dm-dd}
\def \lsim {\:\raisebox{-0.7ex}{$\stackrel{\textstyle<}{\sim}$}\:}
\def \gsim {\:\raisebox{-0.7ex}{$\stackrel{\textstyle>}{\sim}$}\:}
\begin{document}
\thispagestyle{empty}
\begin{flushright}
 March 2011
\end{flushright}
\begin{center}
{\Large\bf
 Determining Ratios of WIMP--Nucleon Cross Sections \\ \vspace{0.2cm}
 from Direct Dark Matter Detection Data}            \\
\vspace{0.7cm}
 {\sc Chung-Lin Shan} \\
\vspace{0.5cm}
 {\it Department of Physics, National Cheng Kung University      \\
      No.~1, University Road,
      Tainan City 70101, Taiwan, R.O.C.}                         \\~\\
 {\it Physics Division,
      National Center for Theoretical Sciences                   \\
      No.~101, Sec.~2, Kuang-Fu Road,
      Hsinchu City 30013, Taiwan, R.O.C.}                        \\~\\
 {\it E-mail:} {\tt clshan@mail.ncku.edu.tw}                     \\
\end{center}
\vspace{1cm}
\begin{abstract}
 Weakly Interacting Massive Particles (WIMPs) are
 one of the leading candidates for Dark Matter.
 So far the usual procedure for constraining
 the WIMP--nucleon cross sections
 in direct Dark Matter detection experiments
 have been to fit the predicted event rate
 based on some model(s) of the Galactic halo
 and of WIMPs to experimental data.
 One has to assume
 whether the spin--independent (SI) or the spin--dependent (SD)
 WIMP--nucleus interaction dominates,
 and results of such data analyses are also expressed
 as functions of the as yet unknown WIMP mass.
 In this article,
 I introduce methods for extracting information on
 the WIMP--nucleon cross sections by considering
 a general combination of the SI and SD interactions.
 {\em Neither} prior knowledge about
 the local density and the velocity distribution of halo WIMPs
 {\em nor} about their mass is needed.
 Assuming that
 an exponential--like shape of the recoil spectrum
 is confirmed from experimental data,
 the required information
 are only the measured recoil energies
 (in low energy ranges)
 and the number of events in the first energy bin
 from two or more experiments.
\end{abstract}
\clearpage
\section{Introduction}
 Astronomical observations and measurements indicate that
 more than 80\% of all matter in the Universe is dark
 (i.e., interacts at most very weakly
  with electromagnetic radiation and ordinary matter).
 The dominant component of this cosmological Dark Matter
 must be due to some yet to be discovered, non--baryonic particles.
 Weakly Interacting Massive Particles (WIMPs) $\chi$
 arising in several extensions of
 the Standard Model of electroweak interactions
 are one of the leading candidates for Dark Matter.
 WIMPs are stable particles
 with masses roughly between 10 GeV and a few TeV
 and interact with ordinary matter only weakly
 (for reviews, see Refs.~\cite{SUSYDM96, Bertone05}).

 Currently,
 the most promising method to detect different WIMP candidates
 is the direct detection of the recoil energy
 deposited in a low--background underground detector
 by elastic scattering of ambient WIMPs off target nuclei
 \cite{Smith90, Lewin96}.
 The basic expression for the differential event rate
 for elastic WIMP--nucleus scattering is given by \cite{SUSYDM96}:
\beq
   \dRdQ
 = \afrac{\rho_0 \sigma_0}{2 \mchi \mrN^2}
   \FQ \int_{\vmin}^{\vmax} \bfrac{f_1(v)}{v} dv
\~.
\label{eqn:dRdQ}
\eeq
 Here $R$ is the direct detection event rate,
 i.e., the number of events
 per unit time and unit mass of detector material,
 $Q$ is the energy deposited in the detector,
 $\rho_0$ is the WIMP density near the Earth,
 $\sigma_0$ is the total cross section
 ignoring the form factor suppression and
 $F(Q)$ is the elastic nuclear form factor,
 $f_1(v)$ is the one--dimensional velocity distribution function
 of the WIMPs impinging on the detector,
 $v$ is the absolute value of the WIMP velocity
 in the laboratory frame.
 The reduced mass $\mrN$ is defined by
\beq
        \mrN
 \equiv \frac{\mchi \mN}{\mchi + \mN}
\~,
\label{eqn:mrN}
\eeq
 where $\mchi$ is the WIMP mass and
 $\mN$ that of the target nucleus.
 Finally,
 $\vmin$ is the minimal incoming velocity of incident WIMPs
 that can deposit the energy $Q$ in the detector:
\beq
   \vmin
 = \alpha \sqrt{Q}
\~,
\label{eqn:vmin}
\eeq
 with the transformation constant
\beq
        \alpha
 \equiv \sfrac{\mN}{2 \mrN^2}
\~,
\label{eqn:alpha}
\eeq
 and $\vmax$ is the maximal WIMP velocity
 in the Earth's reference frame,
 which is related to
 the escape velocity from our Galaxy
 at the position of the Solar system,
 $\vesc~\gsim~600$ km/s.
\subsection{WIMP--nucleus cross section}
 The total WIMP--nucleus cross section
 $\sigma_0$ in Eq.~(\ref{eqn:dRdQ})
 depends on the nature of WIMP couplings on nucleons.
 Generally,
 for non--relativistic WIMPs,
 one can distinguish spin--independent (SI)
 and spin--dependent (SD) couplings.
\subsubsection{Spin--independent couplings}
 Through e.g., squark and Higgs exchanges with quarks,
 WIMPs could have a ``scalar'' interaction with nuclei%
\footnote{
 Besides of the scalar interaction,
 WIMPs could also have a ``vector'' interaction
 with nuclei
 \cite{SUSYDM96, Bertone05}:
\beq
   \sigma_0^{\rm vector}
 = \afrac{1}{64 \pi} \mrN^2
   \bBig{2 Z b_{\rm p} + (A - Z) b_{\rm n}}^2
\~,
\label{eqn:sigma0_vector}
\eeq
 where $b_{\rm (p, n)}$ are the effective
 vector couplings on protons and on neutrons,
 respectively.
 However,
 for Majorana WIMPs ($\chi = \bar{\chi}$),
 e.g., the lightest neutralino in supersymmetric models,
 there is no such vector interaction.
}.
 The total cross section for the SI scalar interaction
 can be expressed as \cite{SUSYDM96, Bertone05}
\beq
   \sigmaSI
 = \afrac{4}{\pi} \mrN^2 \bBig{Z f_{\rm p} + (A - Z) f_{\rm n}}^2
\~.
\label{eqn:sigma0_scalar}
\eeq
 Here $\mrN$ is the reduced mass defined in Eq.~(\ref{eqn:mrN}),
 $Z$ is the atomic number of the target nucleus,
 i.e., the number of protons,
 $A$ is the atomic mass number,
 $A-Z$ is then the number of neutrons,
 $f_{\rm (p, n)}$ are the effective
 scalar couplings of WIMPs on protons p and on neutrons n,
 respectively.
 Here we have to sum over the couplings
 on each nucleon before squaring
 because the wavelength associated with the momentum transfer
 is comparable to or larger than the size of the nucleus,
 the so--called ``coherence effect''.

 In addition,
 for the lightest supersymmetric neutralino,
 and for all WIMPs which interact primarily through Higgs exchange,
 the scalar couplings are approximately the same
 on protons and on neutrons \cite{Cotta09}:
\beq
        f_{\rm n}
 \simeq f_{\rm p}
\~.
\label{eqn:fp/n}
\eeq
 The ``pointlike'' cross section $\sigmaSI$
 in Eq.~(\ref{eqn:sigma0_scalar}) can thus be written as
\beq
        \sigmaSI
 \simeq \afrac{4}{\pi} \mrN^2 A^2 |f_{\rm p}|^2
 =      A^2 \afrac{\mrN}{\mrp}^2 \sigmapSI
\~,
\label{eqn:sigma0SI}
\eeq
 where $\mrp$ is the reduced mass
 of the WIMP mass $\mchi$ and the proton mass $m_{\rm p}$,
 and
\beq
   \sigmapSI
 = \afrac{4}{\pi} \mrp^2 |f_{\rm p}|^2
\label{eqn:sigmapSI}
\eeq
 is the SI WIMP--nucleon cross section.
 The tiny mass difference between a proton and a neutron
 has been neglected.
\subsubsection{Spin--dependent couplings}
 Through e.g., squark and Z boson exchanges with quarks,
 WIMPs could also couple to the spin of target nuclei,
 an ``axial--vector'' (spin--spin) interaction.
 The SD WIMP--nucleus cross section
 can be expressed as \cite{SUSYDM96, Bertone05}:
\beq
   \sigmaSD
 = \afrac{32}{\pi} G_F^2 \~ \mrN^2
   \afrac{J + 1}{J} \bBig{\Srmp \armp + \Srmn \armn}^2
\~.
\label{eqn:sigma0SD}
\eeq
 Here $G_F$ is the Fermi constant,
 $J$ is the total spin of the target nucleus,
 $\expv{S_{\rm (p, n)}}$ are the expectation values of
 the proton and neutron group spins,
 and $a_{\rm (p, n)}$ are the effective SD WIMP couplings
 on protons and on neutrons.

 For the SD WIMP--nucleus interaction,
 it is usually assumed that
 only unpaired nucleons contribute significantly
 to the total cross section,
 as the spins of the nucleons in a nucleus
 are systematically anti--aligned%
\footnote{
 However,
 more detailed nuclear spin structure calculations show that
 the even group of nucleons has sometimes
 also a non--negligible spin
 (see Table 1 and
  e.g., data given in Refs.~\cite{SUSYDM96, Tovey00, Giuliani05}).
 Hence,
 due to the neglect of the contribution
 from the even group of the target nucleons,
 the (exclusion limit of the) WIMP--nucleon cross sections
 could be {\em overestimated}.
}.
 Under the ``odd--group'' assumption,
 the SD WIMP--nucleus cross section can be reduced to
\beq
   \sigmaSD
 = \afrac{32}{\pi} G_F^2 \~ \mrN^2
   \afrac{J + 1}{J} \expv{S_{\rm (p, n)}}^2 |a_{\rm (p, n)}|^2
\~.
\label{eqn:sigma0SD_odd}
\eeq
 Since
 for a proton or a neutron
 $J = \frac{1}{2}$ and $\Srmp$ or $\Srmn = \frac{1}{2}$,
 the SD WIMP cross section on protons or on neutrons
 can be given as
\beq
   \sigma_{\chi {\rm (p, n)}}^{\rm SD}
 = \afrac{24}{\pi} G_F^2 \~ m_{\rm r, (p, n)}^2 |a_{\rm (p, n)}|^2
\~.
\label{eqn:sigmap/nSD}
\eeq
\begin{table}[t!]
\small
\begin{center}
\renewcommand{\arraystretch}{1.35}
\begin{tabular}{|| c   c   c   c   c   c   c   c ||}
\hline
\hline
 \makebox[1.3cm][c]{Isotope}        &
 \makebox[0.9cm][c]{$Z$}            & \makebox[0.9cm][c]{$J$}     &
 \makebox[1.5cm][c]{$\Srmp$}        & \makebox[1.5cm][c]{$\Srmn$} &
 \makebox[1.8cm][c]{$-\Srmp/\Srmn$} & \makebox[1.8cm][c]{$\Srmn/\Srmp$} &
 \makebox[3.7cm][c]{Natural abundance (\%)} \\
\hline
\hline
 $\rmXA{F}{19}$   &  9 & 1/2 &                   0.441  & \hspace{-1.8ex}$-$0.109 &
      4.05  &  $-$0.25   &       100   \\
\hline
 $\rmXA{Na}{23}$  & 11 & 3/2 &                   0.248  &                   0.020 &
  $-$12.40  &     0.08   &       100   \\
\hline
 $\rmXA{Cl}{35}$  & 17 & 3/2 & \hspace{-1.8ex}$-$0.059  & \hspace{-1.8ex}$-$0.011 &
   $-$5.36  &     0.19   &        76   \\
\hline
 $\rmXA{Cl}{37}$  & 17 & 3/2 & \hspace{-1.8ex}$-$0.058  &                   0.050 &
      1.16  &  $-$0.86   &        24   \\
\hline
 $\rmXA{Ge}{73}$  & 32 & 9/2 &                   0.030  &                   0.378 &
   $-$0.08  &    12.6    &  7.8 / 86 (HDMS) \cite{Bednyakov08a}\\
\hline
 $\rmXA{I}{127}$  & 53 & 5/2 &                   0.309  &                   0.075 &
   $-$4.12  &     0.24   &       100   \\
\hline
 $\rmXA{Xe}{129}$ & 54 &  1/2 &                  0.028  &                   0.359 &
   $-$0.08  &    12.8    &        26   \\
\hline
 $\rmXA{Xe}{131}$ & 54 &  3/2 & \hspace{-1.8ex}$-$0.009 & \hspace{-1.8ex}$-$0.227 &
   $-$0.04  &    25.2    &         21   \\
\hline
\hline
\end{tabular}
\end{center}
\caption{
 List of the relevant spin values of the nuclei
 used for simulations presented in this paper.
 More details can be found in
 e.g., Refs.~\cite{SUSYDM96, Tovey00, Giuliani05, Girard05}.
}
\end{table}

 Moreover,
 once the upper and/or lower limits on
 the WIMP--nucleon cross sections
 have been estimated by Eq.~(\ref{eqn:sigmap/nSD}),
 it has been shown that,
 for a {\em particular WIMP mass},
 one can use the following inequality
 to give constraints on
 the SD WIMP--nucleon couplings on the $\armp - \armn$ plane
 \cite{Tovey00, Giuliani04, Girard05}:
\beq
     \abrac{    \frac{\armp}{\sqrt{\sigma_{\chi {\rm p}}^{\rm SD, upper}}}
            \pm \frac{\armn}{\sqrt{\sigma_{\chi {\rm n}}^{\rm SD, upper}}}}^2
 \le \frac{\pi}{24 G_F^2 m_{\rm r, p}^2}
 \le \abrac{    \frac{\armp}{\sqrt{\sigma_{\chi {\rm p}}^{\rm SD, lower}}}
            \pm \frac{\armn}{\sqrt{\sigma_{\chi {\rm n}}^{\rm SD, lower}}}}^2
\~.
\label{eqn:ap/n_contour}
\eeq
 Here $\sigma_{\chi ({\rm p, n})}^{\rm SD, (upper, lower)}$ are
 the upper/lower limits on
 the SD WIMP--proton/neutron cross sections,
 respectively,
 and the ``$\pm$'' sign in the parenthesis
 is the same as that of the $\Srmn/\Srmp$ ratio.
 So far the best constraint on
 the SD WIMP--proton coupling comes from
 the NAIAD \cite{Alner05},
 KIMS \cite{HSLee07},
 SIMPLE \cite{Felizardo10},
 PICASSO \cite{Piro10},
 and COUPP \cite{Behnke10} experiments:
 $|\armp|~\lsim~0.4$ (for a WIMP mass of 50 GeV$/c^2$)
 \cite{Felizardo10},
 whereas the best one on
 the SD WIMP--neutron coupling comes from
 the CDMS-II \cite{Akerib05c},
 XENON10 \cite{Angle08},
 and ZEPLIN-III \cite{Lebedenko09} experiments:
 $|\armn|~\lsim~0.2$ (for a WIMP mass of 50 GeV$/c^2$)
 \cite{Felizardo10}.
 On the other hand,
 the relative strength of two couplings for neutralino WIMPs
 has been calculated as \cite{Bednyakov04}%
\footnote{
 However,
 Ellis {\it et al.} have shown that,
 in different theoretical scenarios,
 $\armn$ could also be slightly greater than $\armp$
 \cite{Ellis08, Cotta09}.
}
\beq
   0.55
 < \vfrac{\armn}{\armp}
 < 0.8
\~.
\label{eqn:ranap_range_th}
\eeq
 Remind that
 the above conventional data analyses are
 independent of models of WIMP--nucleon couplings,
 but they {\em do depend} on the model of the Galactic halo
 through the use of the local WIMP density, $\rho_0$,
 and the velocity distribution of incident WIMPs, $f_1(v)$.
 Additionally,
 the results depend also strongly
 on the as yet unknown WIMP mass
 (see e.g., Refs.~\cite{Giuliani04, Giuliani05, Girard05}).
\subsubsection{Comparison of the SI and SD interactions}
 As discussed above,
 WIMPs could have both SI and SD interactions
 with target nuclei.
 Thus the WIMP--nucleus cross section
 $\sigma_0$ in Eq.~(\ref{eqn:dRdQ})
 should be a combination of the SI cross section
 $\sigmaSI$ in Eq.~(\ref{eqn:sigma0_scalar})
 and the SD cross section
 $\sigmaSD$ in Eq.~(\ref{eqn:sigma0SD}).
 However,
 due to the coherence effect with the entire nucleus
 shown in Eq.~(\ref{eqn:sigma0SI}),
 the cross section for scalar interaction
 scales approximately as the square of
 the atomic mass number of the target nucleus.
 Hence,
 in most supersymmetric models,
 the SI cross section for nuclei with $A~\gsim~30$ dominates
 over the SD one \cite{SUSYDM96, Bertone05}.
 Nevertheless,
 as discussed in Refs.~\cite{Bertone07, Barger08, Belanger08},
 in Universal Extra Dimension (UED) models,
 the SD WIMP interaction with nucleus is less suppressed
 and could be compatible or even larger than the SI one.
\subsection{Nuclear form factor}
\subsubsection{For the spin--independent cross section}
 For the SI cross section,
 there are some analytic forms
 for the elastic nuclear form factor.
 The simplest one is the exponential form factor,
 first introduced by Ahlen {\it et al.} \cite{Ahlen87}
 and Freese {\it et al.} \cite{Freese88}:
\beq
   F_{\rm SI}^2(Q)
 = e^{-Q / Q_0}
\~.
\label{eqn:FQ_ex}
\eeq
 Here $Q$ is the recoil energy
 transferred from the incident WIMP to the target nucleus,
\beq
   Q_0
 = \frac{1.5}{\mN R_0^2}
\label{eqn:Q0}
\eeq
 is the nuclear coherence energy and
\beq
   R_0
 = \bbrac{0.3 + 0.91 \afrac{\mN}{\rm GeV}^{1/3}}~{\rm fm}
\label{eqn:R0}
\eeq
 is the radius of the nucleus.
 The exponential form factor implies
 a Gaussian form of the radial density profile of the nucleus.
 This Gaussian density profile is simple,
 but not very realistic.
 Engel has therefore suggested
 to use the following one \cite{Engel91},
 which derives from the nuclear density profile
 obtained by convolving a constant nuclear density
 with a gaussian one \cite{Helm56},
 and is similar to the numerical form factor
 derived from the Woods--Saxon nuclear density profile
 \cite{SUSYDM96, Bertone05},
\beq
   F_{\rm SI}^2(Q)
 = \bfrac{3 j_1(q R_1)}{q R_1}^2 e^{-(q s)^2}
\~.
\label{eqn:FQ_WS}
\eeq
 Here $j_1(x)$ is a spherical Bessel function,
\beq
   q
 = \sqrt{2 m_{\rm N} Q}
\label{eqn:qq}
\eeq
 is the transferred 3-momentum,
\beq
   R_1
 = \sqrt{R_A^2 - 5 s^2}
\label{eqn:R1}
\eeq
 is the effective nuclear radius%
\footnote{
 In the literature,
 another often used analytic form for $R_1$
 has been given as \cite{Helm56, Lewin96}
\beq
   R_1
 = \sqrt{R_A^2 + {\T \afrac{7}{3}} \pi^2 r_0^2 - 5 s^2}
\~,
\label{eqn:R1_Helm}
\eeq
 where
\beq
        R_A
 \simeq \abig{1.23 \~ A^{1/3} - 0.6}~{\rm fm},
        ~~~~~~~~~~~~~~ 
        r_0
 \simeq 0.52~{\rm fm},
        ~~~~~~~~~~~~~~ 
        s
 \simeq 0.9~{\rm fm}.
\label{eqn:RA_r0_ss_Helm}
\eeq
}
 with%
\footnote{
 For $R_1$ given by Eq.~(\ref{eqn:R1})
 with $s \simeq 1$ fm,
 another analytic form for $R_A$
 has also been given \cite{Eder68, Lewin96}:
\beq
        R_A
 \simeq \abig{1.15 \~ A^{1/3} + 0.39}~{\rm fm}.
\label{eqn:RA_Eder}
\eeq
}
\beq
        R_A
 \simeq 1.2 \~ A^{1/3}~{\rm fm},
\label{eqn:RA}
\eeq
 and
\beq
        s
 \simeq 1~{\rm fm}
\label{eqn:ss}
\eeq
 is the nuclear skin thickness.
\subsubsection{For the spin--dependent cross section}
 For the SD cross section,
 the form factor is different from nucleus to nucleus
 and no simple analytic form
 can provide a very good approximation.
 Generally,
 the form factor for the SD cross section
 can be expressed as \cite{Lewin96, SUSYDM96}
\beq
   F_{\rm SD}^2(Q)
 = \frac{S(q)}{S(0)}
\~.
\label{eqn:FQ_SD_SS}
\eeq
 Here the ``spin structure'' function $S(q)$
 depends generally on the SD WIMP--nucleon couplings:
\beq
   S(q)
 = a_0^2 S_{00}(q) + a_1^2 S_{11}(q) + a_0 a_1 S_{01}(q)
\~,
\label{eqn:Sq}
\eeq
 with the isoscalar and isovector coefficients:
\beq
   a_0
 = \armp + \armn
\~,
   ~~~~~~~~~~~~~~~~~~~~ 
   a_1
 = \armp - \armn
\~,
\label{eqn:a0_a1}
\eeq
 and $S_{00}$, $S_{11}$, and $S_{01}$ are
 the isoscalar, isovector and interference contributions to $S(q)$,
 respectively.

 However,
 Klapdor-Kleingrothaus {\it et al.}
 have used the following form factor
 for the SD cross section \cite{Klapdor05},
 introduced by Lewin and Smith
 with the so--called ``thin--shell'' approximation
 \cite{Lewin96}:
\beqn
    F_{\rm SD}^2(Q)
 \= \cleft{\renewcommand{\arraystretch}{1.5}
           \begin{array}{l l l}
            j_0^2(q R_1)                      \~, & ~~~~~~~~ &
            {\rm for}~q R_1 \le 2.55~{\rm or}~q R_1 \ge 4.5 \~, \\ 
            {\rm const.} \simeq 0.047         \~, &          &
            {\rm for}~2.5 5 \le q R_1 \le 4.5 \~.  
           \end{array}}
\label{eqn:FQ_TS}
\eeqn
\subsubsection{Zero momentum transfer approximation}
 For our simulations presented in this article,
 we will use the form factors
 given by Eqs.~(\ref{eqn:FQ_WS}) and (\ref{eqn:FQ_TS})
 for the SI and SD cross sections,
 respectively.
 However,
 it will be seen later that,
 since one would only have to estimate values of the form factors
 at the lowest energy ranges
 (\mbox{$\lsim$ 20 keV} for some currently running
  and projected experiments),
 we could practically use the
 ``zero momentum transfer'' approximation:
\beq
        F^2(Q \simeq 0)
 \simeq 1
\label{eqn:FQ_approx}
\eeq
 in the methods introduced in this article.
\subsection{Motivation}
 So far the usual procedure for estimating
 the (exclusion limits of the) WIMP--nucleon cross sections
 in direct Dark Matter detection experiments
 have been to fit the predicted event rate,
 $dR/dQ$ in Eq.~(\ref{eqn:dRdQ}),
 based on some model(s) of the Galactic halo from cosmology
 and of WIMPs from particle physics
 to experimental data.
 Meanwhile,
 one has to assume whether
 the SI or the SD WIMP--nucleus interaction dominates.
 However,
 WIMPs should in general have
 both interactions with target nuclei.
 Moreover,
 as mentioned above,
 although in most models with neutralino WIMPs
 as the best motivated candidate for Dark Matter,
 the theoretical predicted SI WIMP--nucleus cross section
 should be (much) larger than the SD one \cite{Cotta09},
 Bertone {\it et al.} have shown that
 another Dark Matter candidate,
 the lightest Kaluza--Klein particle (LKP)
 arising in the Universal Extra Dimension (UED) models,
 has a relatively larger SD cross section,
 or, equivalently,
 a larger $\sigma_{\chi ({\rm p, n})}^{\rm SD}$ to $\sigmapSI$ ratio
 \cite{Bertone07}.
 Hence,
 for determining the nature of Dark Matter particles
 and distinguishing them
 between e.g., the lightest neutralino in supersymmetric models
 and the lightest Kaluza--Klein particles
 in models with Universal Extra Dimensions,
 estimates of both SI and SD cross sections,
 or, at least an estimate of the ratio
 between these two cross sections,
 in direct Dark Matter detection experiments
 is essential.

 On the other hand,
 as shown in our earlier work
 \cite{DMDDmchi-SUSY07, DMDDmchi} that
 one can determine the WIMP mass
 with direct Dark Matter detection experiments
 {\em without} a prior knowledge of the WIMP--nucleus cross section
 {\em nor} assumptions about the local density
 and the velocity distribution function of halo WIMPs.
 It is therefore important to investigate methods
 for, conversely, extracting information on
 the WIMP--nucleon cross sections from experimental data
 {\em without} knowing the WIMP mass.

 The remainder of this article is organized as follows.
 In Secs.~2 and 3
 I will show how to determine ratios of
 WIMP--nucleon couplings/cross sections
 once positive signals have been observed.
 Both the case that
 the SD WIMP interaction dominates (in Sec.~2)
 and that of a general combination of
 the SI and SD cross sections (in Sec.~3)
 will be considered.
 In Sec.~4
 I will extend the data analysis procedure
 to the estimates of ratios between the SI WIMP scalar/vector couplings
 on protons and on neutrons.
 I conclude in Sec.~5.
 Some technical details for the data analysis
 will be given in an appendix.
\section{Only a dominant SD WIMP--nucleus cross section}
 In this section
 I consider at first the case
 that the SD WIMP--nucleus interaction dominates
 over the SI one
 and derive the expression
 for determining the ratio between
 two SD WIMP--nucleon couplings.
\subsection{General expression}
 By using a time--averaged recoil spectrum,
 and assuming that no directional information exists,
 the normalized one--dimensional
 velocity distribution function of halo WIMPs, $f_1(v)$,
 has been solved from Eq.~(\ref{eqn:dRdQ}) analytically \cite{DMDDf1v}
 and,
 consequently,
 its generalized moments can be estimated by
 \cite{DMDDf1v, DMDDmchi}%
\footnote{
 Here we have implicitly assumed that
 $\Qmax$ is so large that
 terms involving $- 2 \Qmax^{(n+1)/2} r(\Qmax) / F^2(\Qmax)$
 are negligible.
 Due to sizable contributions
 from large recoil energies \cite{DMDDf1v},
 this is not necessarily true, 
 especially for some not--very--high $\Qmax$
 in the experimental reality,
 and/or heavy detector targets,
 and/or heavy WIMPs.
 Nevertheless,
 we will show in this and the next sections that,
 since we use only $n = -1$, 1, and 2,
 Eq.~(\ref{eqn:moments}) can still be used
 for the determinations of the ratios
 between different WIMP--nucleon couplings/cross sections.
 Moreover,
 considering the large statistical uncertainties
 due to (very) few events in the highest energy ranges,
 this should practically be a good approximation.
}
\beqn
    \expv{v^n}(v(\Qmin), v(\Qmax))
 \= \int_{v(\Qmin)}^{v(\Qmax)} v^n f_1(v) \~ dv
    \non\\
 \= \alpha^n
    \bfrac{2 \Qmin^{(n+1)/2} r(\Qmin) / \FQmin + (n+1) I_n(\Qmin, \Qmax)}
          {2 \Qmin^{   1 /2} r(\Qmin) / \FQmin +       I_0(\Qmin, \Qmax)}
\~.
\label{eqn:moments}
\eeqn
 Here $v(Q) = \alpha \sqrt{Q}$,
 $Q_{\rm (min, max)}$ are
 the experimental minimal and maximal
 cut--off energies of the data set,
 respectively,
\beq
        r(\Qmin)
 \equiv \adRdQ_{{\rm expt}, \~ Q = \Qmin}
\label{eqn:rmin}
\eeq
 is an estimated value of
 the {\em measured} recoil spectrum $(dR/dQ)_{\rm expt}$
 ({\em before} normalized by
  an experimental exposure $\cal E$)
 at $Q = \Qmin$,
 and $I_n(\Qmin, \Qmax)$ can be estimated through the sum:
\beq
   I_n(\Qmin, \Qmax)
 = \sum_{a = 1}^{N_{\rm tot}} \frac{Q_a^{(n-1)/2}}{F^2(Q_a)}
\~,
\label{eqn:In_sum}
\eeq
 where the sum runs over all events in the data set
 that satisfy $Q_a \in [\Qmin, \Qmax]$
 and $N_{\rm tot}$ is the number of such events.

 Now,
 since the integral on the right--hand side of Eq.~(\ref{eqn:dRdQ})
 is just the minus--first moment of
 the velocity distribution function, $\expv{v^{-1}}$,
 which can be estimated by Eq.~(\ref{eqn:moments}),
 by setting $Q = \Qmin$ and
 using the definition (\ref{eqn:alpha}) of $\alpha$,
 one can obtain straightforwardly that
\beq
   \rho_0 \sigma_0
 = \afrac{1}{\calE}
   \mchi \mrN \sfrac{\mN}{2}
   \bbrac{\frac{2 \Qmin^{1/2} r(\Qmin)}{\FQmin} + I_0}
\~.
\label{eqn:rho_sigma}
\eeq
 Then,
 in order to avoid the uncertainty of $\rho_0$
 (of a factor of $\sim 2$ \cite{SUSYDM96}),
 one can combine two experimental data sets
 with different target nuclei, $X$ and $Y$,
 to eliminate $\rho_0$ in Eq.~(\ref{eqn:rho_sigma})
 and thus obtain the following expression
 for the ratio between the WIMP cross section
 on nuclei $X$ and $Y$:
\beq
   \frac{\sigmaX}{\sigmaY}
 = \frac{\mrX}{\mrY} \sfrac{\mX}{\mY}
   \afrac{\calR_{\sigma, X}}{\calR_{\sigma, Y}}
\~,
\label{eqn:rsigmaXY}
\eeq
 where $m_{{\rm r}, (X, Y)}$ are the reduced masses
 of the WIMP mass and the masses of target nucleus, $m_{(X, Y)}$,
 and I have defined
\beq
        \calR_{\sigma, X}
 \equiv \frac{1}{\calE_X}
        \bbrac{\frac{2 \QminX^{1/2} r_X(\QminX)}{\FQminX} + \IzX}
\~,
\label{eqn:RsigmaX_min}
\eeq
 and similar for $\calR_{\sigma, Y}$;
 $F_{(X, Y)}(Q)$ here
 are the form factors of the nucleus $X$ and $Y$,
 $r_{(X, Y)}(Q_{{\rm min}, (X, Y)})$
 refer to the counting rates for the target $X$ and $Y$
 at the respective lowest recoil energies included in the analysis,
 and $\calE_{(X, Y)}$ are the experimental exposures
 with the target $X$ and $Y$.
 The emphasize here is that
 Eq.~(\ref{eqn:rsigmaXY}) can be used
 once {\em positive} signals are observed
 in two (or more) experiments;
 information on the local WIMP density $\rho_0$ and
 on the velocity distribution function of halo WIMPs, $f_1(v)$,
 are {\em not} necessary
\footnote{
 Later we will see that
 {\em nor} information on the WIMP mass $\mchi$ is necessary.
}.

 Substituting the expression (\ref{eqn:sigma0SD})
 for $\sigmaSD$ into Eq.~(\ref{eqn:rsigmaXY})
 and using the definition (\ref{eqn:alpha}) of $\alpha$
 for both target nuclei,
 one can solve the ratio
 between two SD WIMP--nucleon couplings analytically as
 \cite{DMDDidentification-DMDE2009}%
\footnote{
 Note that,
 although the constraints on two SD WIMP--nucleon couplings
 have conventionally been shown in the $\armp - \armn$ plane,
 considering the theoretical expected value
 given in Eq.~(\ref{eqn:ranap_range_th}),
 we use always the $\armn / \armp$ ratio
 in our work.
}
\beq
   \afrac{\armn}{\armp}_{\pm, n}^{\rm SD}
 =-\frac{\SpX \pm \SpY \abrac{\calR_{J, n, X} / \calR_{J, n, Y}} }
        {\SnX \pm \SnY \abrac{\calR_{J, n, X} / \calR_{J, n, Y}} }
\~,
\label{eqn:ranapSD}
\eeq
 for $n \ne 0$.
 Here I have used the relation \cite{DMDDmchi}:
\beq
   \frac{\alpha_X}{\alpha_Y}
 = \frac{\calR_{n, Y}}{\calR_{n, X}}
\~,
\label{eqn:ralphaXY}
\eeq
 and defined
\beq
        \calR_{J, n, X}
 \equiv \bbrac{\Afrac{J_X}{J_X + 1}
               \frac{\calR_{\sigma, X}}{\calR_{n, X}}}^{1/2}
\~,
\label{eqn:RJnX}
\eeq
 with $\calR_{\sigma, X}$ defined in Eq.~(\ref{eqn:RsigmaX_min})
 and
\beq
        \calR_{n, X}
 \equiv \bfrac{2 \QminX^{(n+1)/2} r_X(\QminX) / \FQminX + (n+1) \InX}
              {2 \QminX^{   1 /2} r_X(\QminX) / \FQminX +       \IzX}^{1/n}
\~;
\label{eqn:RnX_min}
\eeq
 $\calR_{J, n, Y}$ and $\calR_{n, Y}$
 can be defined analogously%
\footnote{
 Hereafter,
 without special remark
 all notations defined for the target $X$
 can be defined analogously for the target $Y$
 and occasionally for the target $Z$.
}.
 Note that
 Eq.~(\ref{eqn:ranapSD}) is {\em independent} of the WIMP mass
 and can be used for estimating $\armn / \armp$
 with measured recoil energies directly.

\begin{figure}[t!]
\begin{center}
\includegraphics[width=8.5cm]{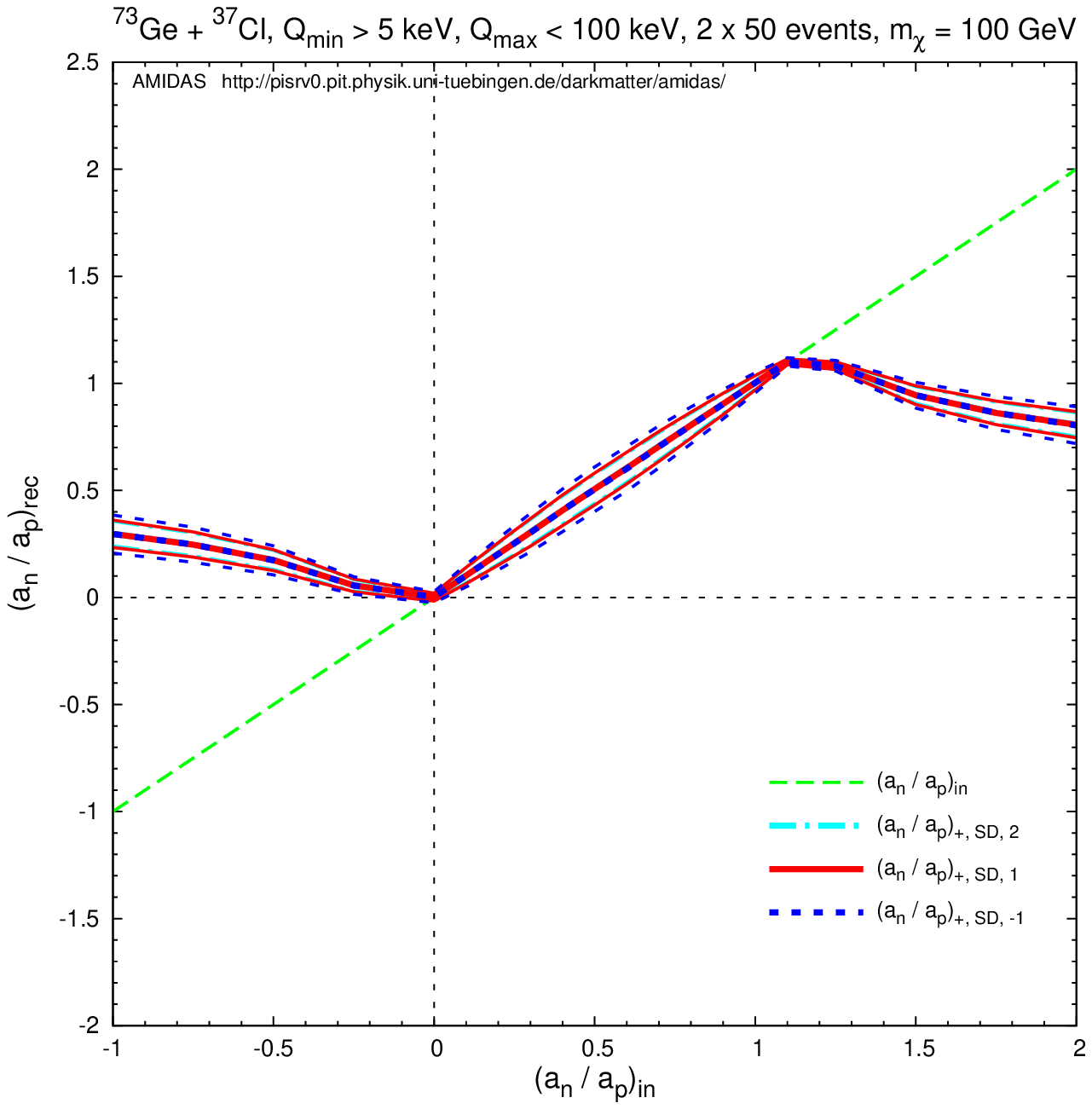}
\includegraphics[width=8.5cm]{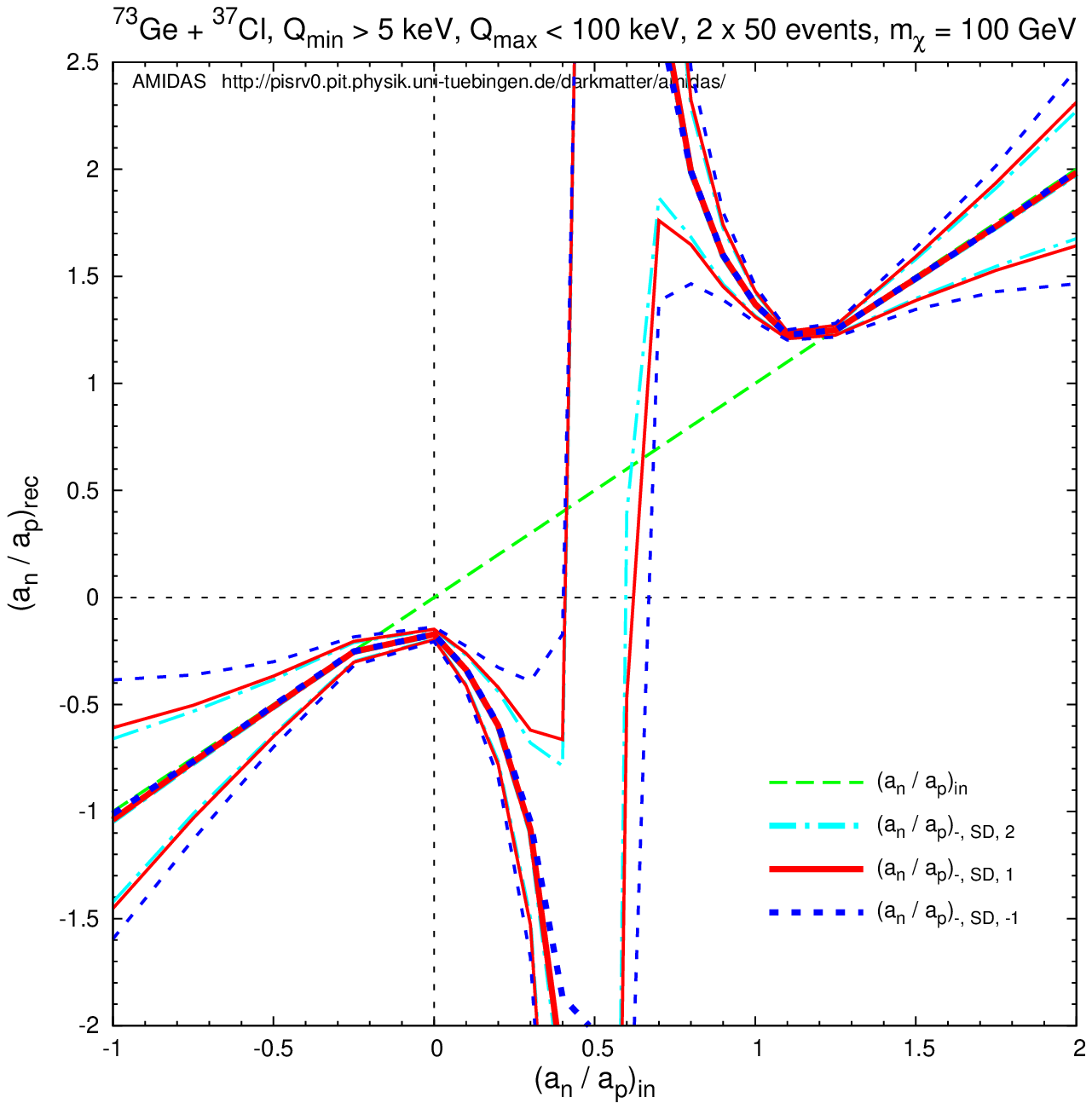}
\vspace{-0.75cm}
\end{center}
\caption{
 The reconstructed $\armn / \armp$ ratios
 estimated by Eq.~(\ref{eqn:ranapSD})
 and the lower and upper bounds of
 their 1$\sigma$ statistical uncertainties
 estimated by Eq.~(\ref{eqn:sigma_ranapSD})
 with $n = -1$ (dashed blue), 1 (solid red),
 and 2 (dash--dotted cyan)
 as functions of the input $\armn / \armp$ ratio.
 Here I show the ``$+$ ($-$)'' solutions
 in the left (right) frames separately.
 The theoretical predicted recoil spectrum for
 the shifted Maxwellian velocity distribution with
 \mbox{$v_0 = 220$ km/s},
 \mbox{$v_{\rm e} = 1.05 \~ v_0$},
 and \mbox{$\vmax = 700$ km/s}
 as well as the nuclear form factor for the SD cross section
 given in Eq.~(\ref{eqn:FQ_TS})
 have been used.
 $\rmXA{Ge}{73}$ and $\rmXA{Cl}{37}$
 have been chosen as two target nuclei.
 Each experiment contains 50 total events on average
 in the energy range between 5 and 100 keV.
 The input WIMP mass has been set as 100 GeV.
 See the text for further details.
}
\label{fig:ranapSD_ranap_pm}
\end{figure}

 Because the couplings in Eq.~(\ref{eqn:sigma0SD}) are squared,
 we have two solutions for $\armn / \armp$ here;
 if exact ``theory'' values for ${\cal R}_{J, n , (X, Y)}$ are taken,
 these solutions coincide for
\beq
   \afrac{\armn}{\armp}_{+, n}^{\rm SD}
 = \afrac{\armn}{\armp}_{-, n}^{\rm SD}
 = \cleft{\renewcommand{\arraystretch}{1}
          \begin{array}{l l l}
           \D -\frac{\SpX}{\SnX}         \~, & ~~~~~~~~ &
           {\rm for}~\calR_{J, n, X} = 0 \~, \\ \\ 
           \D -\frac{\SpY}{\SnY}         \~, &          &
           {\rm for}~\calR_{J, n, Y} = 0 \~,
          \end{array}}
\label{eqn:ranapSD_coin}
\eeq
 which depends only on properties of two used target nuclei
 (see Table 1).
 Moreover,
 it can be found from Eq.~(\ref{eqn:ranapSD}) that
 one of these two solutions has a pole
 at the middle of two intersections,
 which depends simply on the signs of $\SnX$ and $\SnY$:
 since $\calR_{J, n, X}$ and $\calR_{J, n, Y}$ are always positive,
 if both $\SnX$ and $\SnY$ are positive or negative,
 the ``$-$ (minus)'' solution $(\armn / \armp)^{\rm SD}_{-, n}$
 will diverge and
 the ``$+$ (plus)'' solution $(\armn / \armp)^{\rm SD}_{+, n}$
 will be the ``inner'' solution;
 in contrast,
 if the signs of $\SnX$ and $\SnY$ are opposite,
 the ``$-$ (minus)'' solution $(\armn / \armp)^{\rm SD}_{-, n}$
 will be the ``inner'' solution
 (see Figs.~\ref{fig:ranapSD_ranap_pm}).

 By using the standard Gaussian error propagation,
 the statistical uncertainty on
 $\abrac{\armn / \armp}_{\pm, n}^{\rm SD}$
 can be expressed as
\beqn
        \sigma\abrac{\afrac{\armn}{\armp}_{\pm, n}^{\rm SD}}
 \=     \frac{\vBig{\SpY \SnX - \SpX \SnY}}
             {\bBig{\SnX \pm \SnY (\calR_{J, n, X} / \calR_{J, n, Y})}^2}
        \abrac{\frac{1}{2} \cdot \frac{\calR_{J, n, X}}{\calR_{J, n, Y}}}
        \non\\
 \conti ~~ \times
        \cBiggl{  \sum_{i, j = 1}^3
                  \bbrac{  \frac{1}{\calR_{n     , X}} \aPp{\calR_{n     , X}}{c_{i, X}}
                         - \frac{1}{\calR_{\sigma, X}} \aPp{\calR_{\sigma, X}}{c_{i, X}} } }
        \non\\
 \conti ~~~~~~~~~~~~~~~~ \times 
                  \bbrac{  \frac{1}{\calR_{n     , X}} \aPp{\calR_{n     , X}}{c_{j, X}}
                         - \frac{1}{\calR_{\sigma, X}} \aPp{\calR_{\sigma, X}}{c_{j, X}} }
                  {\rm cov}(c_{i, X}, c_{j, X})
        \non\\
 \conti ~~~~~~~~~~~~ 
        \cBiggr{+ (X \lto Y)}^{1/2}
\~.
\label{eqn:sigma_ranapSD}
\eeqn
 Here a short--hand notation for the six quantities
 on which the estimate of $(\armn / \armp)_{\pm, n}^{\rm SD}$ depends
 has been introduced \cite{DMDDmchi}:
\beq
   c_{1, X}
 = I_{n, X}
\~,
   ~~~~~~~~~~~~ 
   c_{2, X}
 = I_{0, X}
\~,
   ~~~~~~~~~~~~ 
   c_{3, X}
 = r_X(\QminX)
\~;
\label{eqn:ciX}
\eeq
 and similarly for the $c_{i, Y}$.
 Estimators for ${\rm cov}(c_i, c_j)$ and
 explicit expressions for the derivatives of
 $\calR_{n, X}$ and $\calR_{\sigma, X}$
 with respect to $c_{i, X}$
 will be given in the appendix.
 Note that
 $\calR_{\sigma, (X, Y)}$ are actually
 independent of $c_{1, (X, Y)} = I_{n, (X, Y)}$,
 for $n \neq 0$.

 In Figs.~\ref{fig:ranapSD_ranap_pm}
 I show the numerical results
 for a target combination of
 $\rmXA{Ge}{73}$ and $\rmXA{Cl}{37}$
 with 5,000 experiments
 based on the Monte Carlo simulation%
\footnote{
 Note that,
 rather than the mean values,
 in this article
 we give always the median values
 of the reconstructed results
 from the simulated experiments.
}.
 The theoretical predicted recoil spectrum for
 the shifted Maxwellian velocity distribution
 \cite{SUSYDM96, Bertone05, DMDDf1v} with
 a Sun's orbital velocity in the Galactic frame $v_0 = 220$ km/s,
 an Earth's velocity in the Galactic frame $v_{\rm e} = 1.05 \~ v_0$,%
\footnote{
 The time dependence of the Earth's velocity
 in the Galactic frame \cite{SUSYDM96, Bertone05}
 has been ignored.
}
 and a maximal cut--off velocity of
 the velocity distribution function $\vmax = 700$ km/s,
 as well as the nuclear form factor given in Eq.~(\ref{eqn:FQ_TS})
 have been used.
 The experimental minimal and maximal cut--off energies
 have been set as \mbox{$\Qmin = 5$ keV} and \mbox{$\Qmax = 100$ keV}
 for both targets.
 Each experiment contains an expected number of 50 total events;
 the actual event number is Poisson--distributed
 around this expectation value.
 The input WIMP mass has been set as 100 GeV.

 As discussed above,
 since $\Srmn_{\rmXA{Ge}{73}}$ and $\Srmn_{\rmXA{Cl}{37}}$
 have the same sign,
 the ``$+$'' solution shown
 in the left frame of Figs.~\ref{fig:ranapSD_ranap_pm}
 is the inner solution
 for the range of interest $0 \le \armn / \armp \le 1$,
 while the ``$-$'' solution shown in the right frame diverges 
 between $- \Srmp_{\rmXA{Ge}{73}} / \Srmn_{\rmXA{Ge}{73}} = -0.08$
 and $- \Srmp_{\rmXA{Cl}{37}} / \Srmn_{\rmXA{Cl}{37}} = 1.16$.
 Note here that,
 for practical use of analyzing {\em real data},
 one might however not be able to make a choice
 from the ``$+$'' and ``$-$'' estimates given by Eq.~(\ref{eqn:ranapSD}),
 especially if they are close to the coincidences,
 e.g., around 1.16 or $-$0.08 here.
 For example,
 for a true $\armn / \armp = 1.1$,
 one will get $(\armn / \armp)_{+}^{\rm SD} \cong 1.1$
 and $(\armn / \armp)_{-}^{\rm SD} \cong 1.25$,
 the same results as for the case
 with a true $\armn / \armp = 1.25$.
\subsection{Reducing statistical uncertainty on
            \boldmath$(\armn / \armp)_{\pm, n}^{\rm SD}$}
 For estimating the statistical uncertainty on
 $(\armn / \armp)_{\pm, n}^{\rm SD}$ by Eq.~(\ref{eqn:sigma_ranapSD}),
 one needs to estimate
 contributions from the counting rate
 at the threshold energy, $r(\Qmin)$,
 from $I_n$ given in Eq.~(\ref{eqn:In_sum}),
 and from the covariance between $r(\Qmin)$ and $I_n$.
 From Eqs.~(\ref{eqn:rmin_eq}), (\ref{eqn:sigma_rmin})
 and (\ref{eqn:cov_rmin_In}) in the appendix,
 one can find a way to reduce
 these statistical uncertainties
 by estimating the counting rate,
 instead of at the experimental minimal cut--off energy,
 at the shifted point $Q_{s, 1}$
 (from the central point of the first bin, $Q_1$):
\beq
   Q_{s, 1}
 = Q_1 + \frac{1}{k_1} \ln\bfrac{\sinh (k_1 b_1 / 2)}{k_1 b_1 / 2}
\~,
\label{eqn:Qs1}
\eeq
 where $k_1$ is the logarithmic slope of
 the reconstructed recoil spectrum
 in the first $Q-$bin and $b_1$ is the bin width.
 Then,
 according to Eq.~(\ref{eqn:rmin_eq}),
 the {\em measured} recoil spectrum
 at $Q = Q_{s, 1}$ can be estimated by
\beq
   r(Q_{s, 1})
 = \afrac{dR}{dQ}_{{\rm expt}, \~ 1, \~ Q = Q_{s, 1}}
 = r_1
 = \frac{N_1}{b_1}
\~,
\label{eqn:rmin_Qs1}
\eeq
 with the statistical uncertainty given as
\beq
   \sigma^2(r(Q_{s, 1}))
 = \sigma^2(r_1)
 = \frac{N_1}{b_1^2}
\~,
\label{eqn:sigma_rmin_Qs1}
\eeq
 where $N_1$ is the event number
 in the first bin.

\begin{figure}[t!]
\begin{center}
\includegraphics[width=8.5cm]{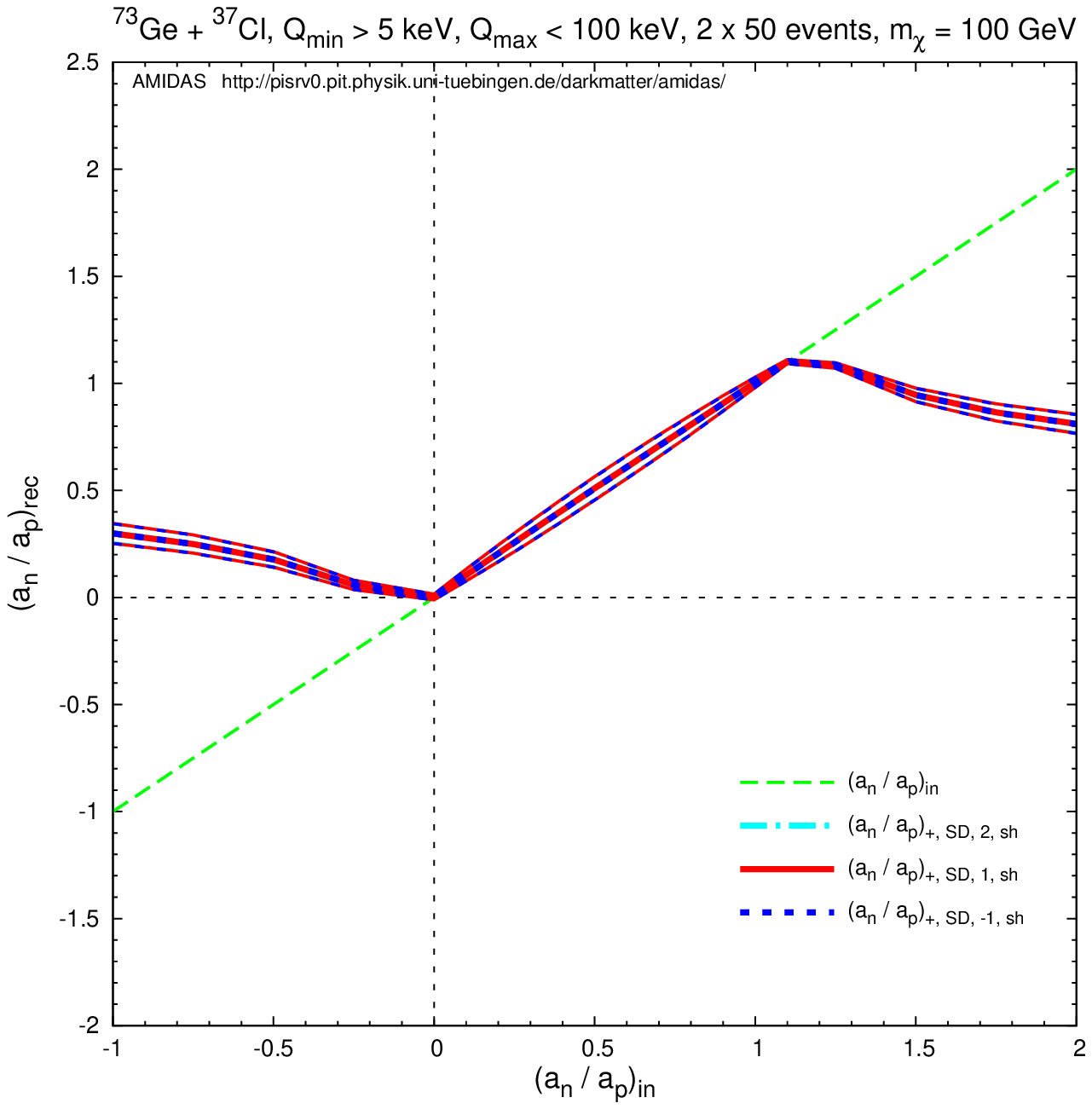}
\includegraphics[width=8.5cm]{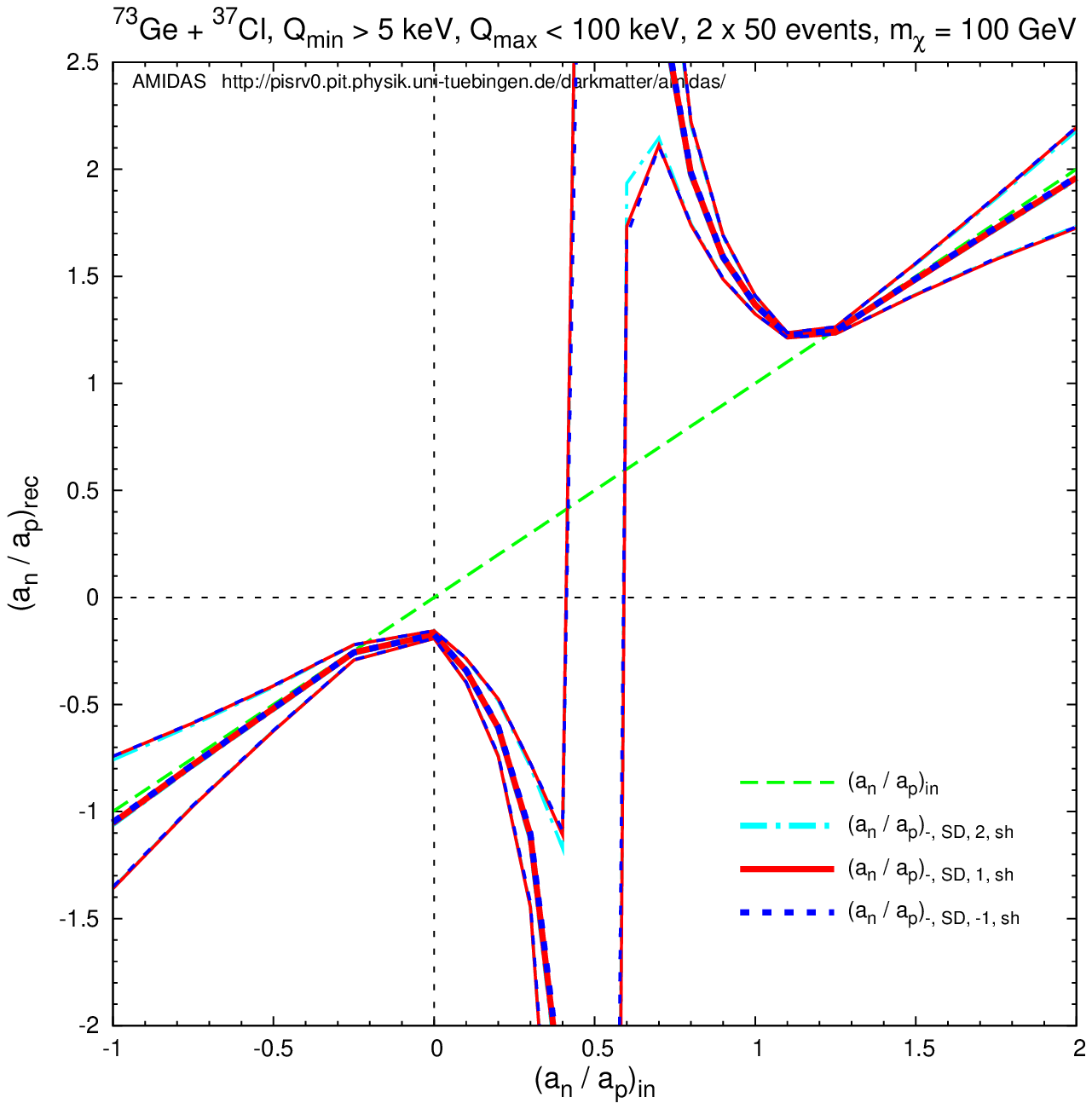}
\vspace{-0.75cm}
\end{center}
\caption{
 As in Figs.~\ref{fig:ranapSD_ranap_pm},
 except that
 we estimate $(\armn / \armp)_{\pm, n}^{\rm SD}$
 with the counting rates
 at the shifted points of the first $Q-$bin,
 $r_{(X, Y)}(Q_{s, 1, (X, Y)}) = r_{(X, Y), 1}$.
}
\label{fig:ranapSD_ranap_sh_pm}
\end{figure}
 In Figs.~\ref{fig:ranapSD_ranap_sh_pm}
 I show the reconstructed $\armn / \armp$ ratios
 and the lower and upper bounds of
 their 1$\sigma$ statistical uncertainties
 with $n = -1$ (dashed blue), 1 (solid red),
 and 2 (dash--dotted cyan)
 estimated by Eq.~(\ref{eqn:ranapSD})
 with the counting rates
 at the shifted points of the first $Q-$bin,
 $r_{(X, Y)}(Q_{s, 1, (X, Y)}) = r_{(X, Y), 1}$
 as functions of the input $\armn / \armp$ ratio%
\footnote{
 Labeled hereafter with an ``sh'' in the subscript.
}.
 It can be seen that
 the statistical uncertainties on $(\armn / \armp)_{\pm, n}^{\rm SD}$
 estimated with different $n$
 (namely with different moments of the WIMP velocity distribution)
 with $r_{(X, Y)}(Q_{s, 1, (X, Y)})$ are clearly reduced and,
 interestingly,
 almost equal.
 Therefore,
 since
\beq
   \calR_{J, -1, X}
 = \bbrac{\afrac{J_X}{J_X + 1}
          \frac{2 \~ r_X(Q_{s, 1, X})}{\calE_X F_X^2(Q_{s, 1, X})}}^{1/2}
\~,
\label{eqn:RJmaX}
\eeq
 one would practically only need events
 in the lowest energy ranges
 ($\sim$ 20 events between 5 and 15 keV in our simulations)
 for estimating $\armn / \armp$.
 Consequently,
 one has to estimate the values of form factors
 only at $Q = Q_{s, 1}$,
 and the zero momentum transfer approximation
 $F^2(Q \simeq 0)) \simeq 1$ can be used.
 In fact,
 our simulation shows that
 a relatively higher threshold energy
 (\mbox{$\Qmin \sim$ 10 keV} and \mbox{$Q_{s, 1} \sim$ 14 keV})
 should not affect the reconstruction of $\armn / \armp$ significantly,
 especially for the first approximation
 with pretty few events and thus a large statistical uncertainty.

\begin{figure}[t!]
\begin{center}
\includegraphics[width=8.5cm]{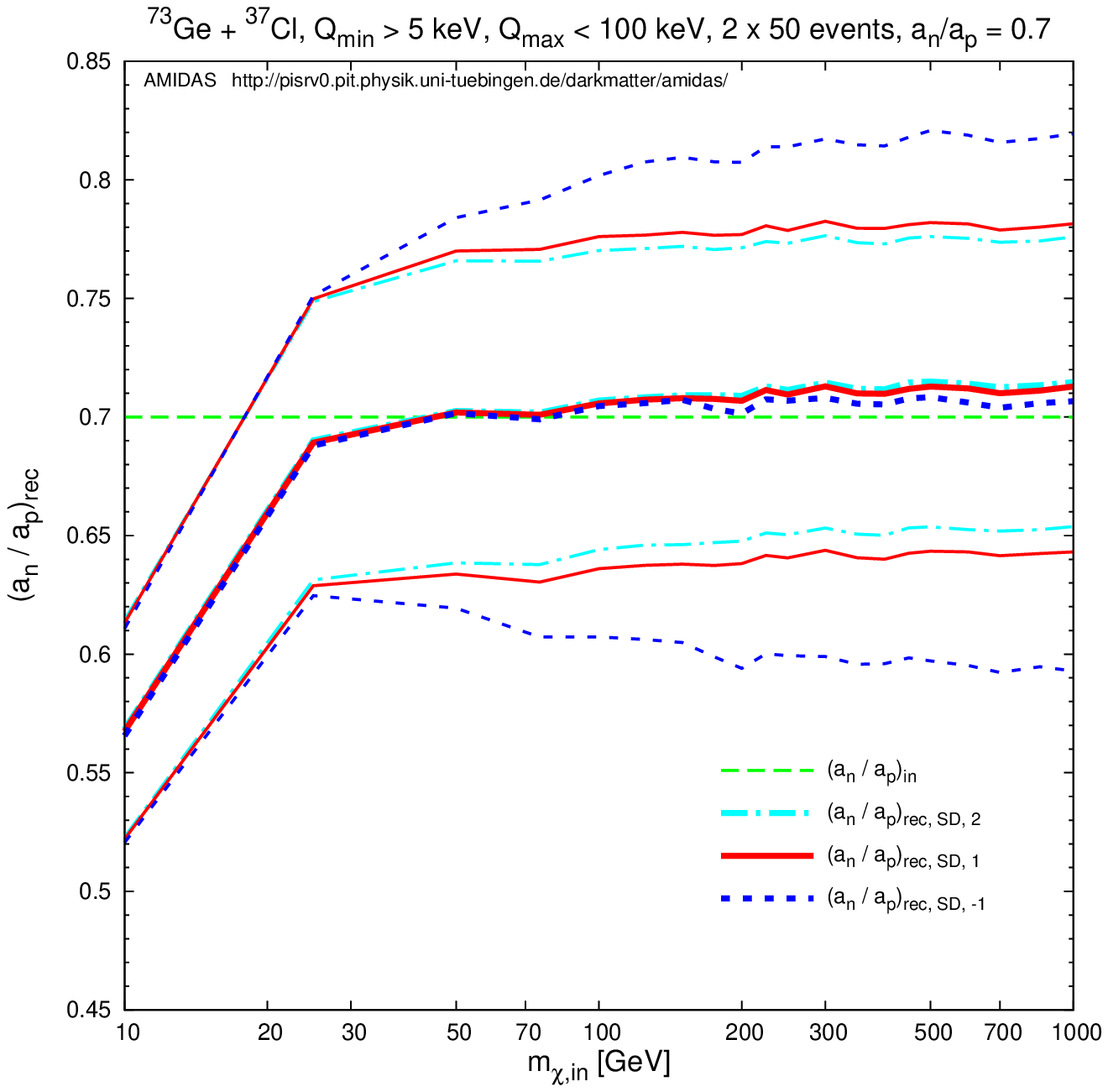}
\includegraphics[width=8.5cm]{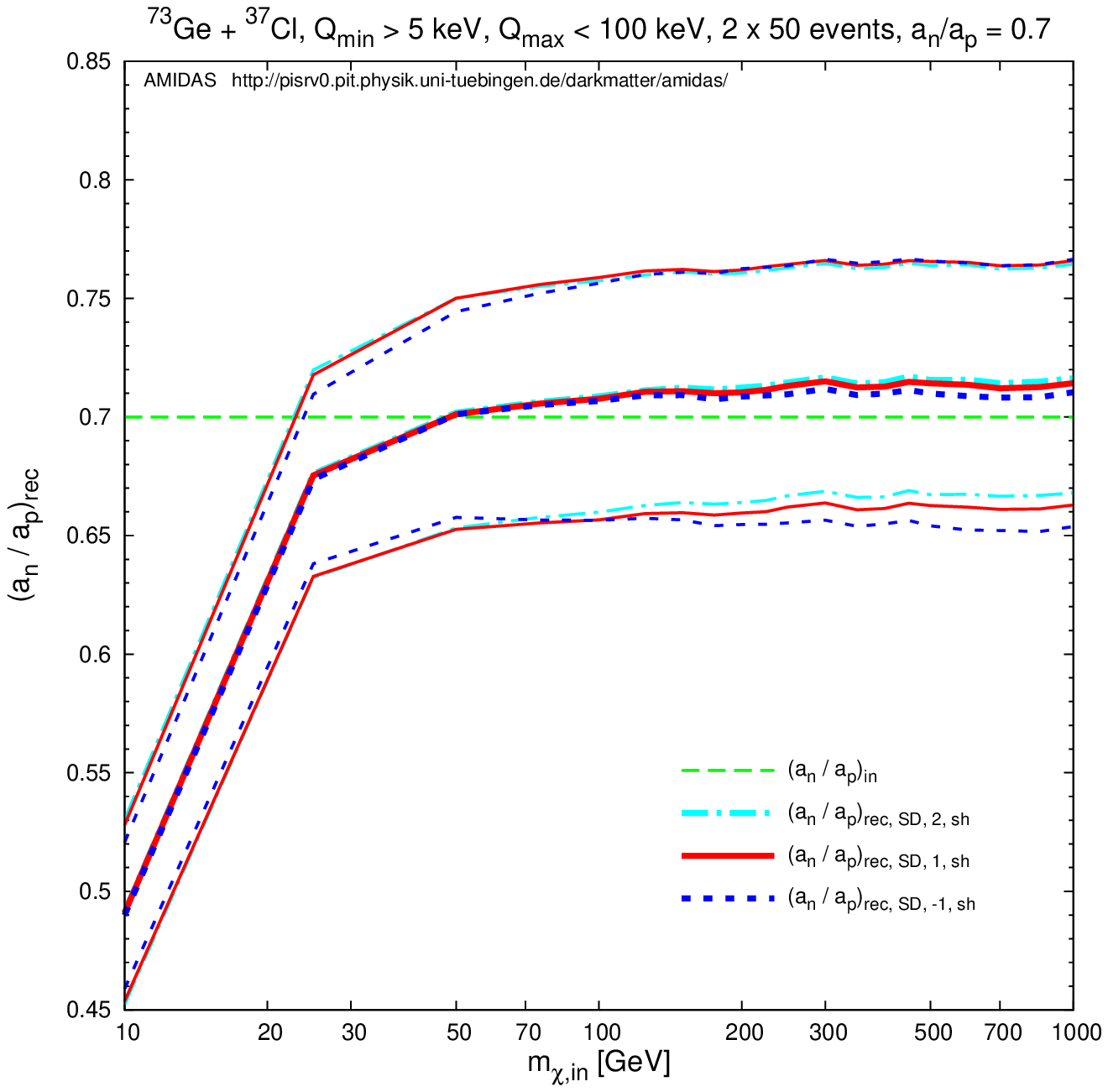}
\vspace{-0.75cm}
\end{center}
\caption{
 The reconstructed $\armn / \armp$ ratios
 estimated by Eq.~(\ref{eqn:ranapSD})
 and the lower and upper bounds of
 their 1$\sigma$ statistical uncertainties
 with $n = -1$ (dashed blue), 1 (solid red),
 and 2 (dash--dotted cyan)
 as functions of the {\em input WIMP mass} $\mchi$.
 Here we estimate
 with $r_{(X, Y)}(Q_{{\rm min}, (X, Y)})$ (left)
 and $r_{(X, Y)}(Q_{s, 1, (X, Y)})$ (right).
 The input $\armn / \armp$ ratio
 has been set as 0.7.
 The other parameters and notations
 are as in Figs.~\ref{fig:ranapSD_ranap_pm}
 and \ref{fig:ranapSD_ranap_sh_pm}.
}
\label{fig:ranapSD_mchi_rec}
\end{figure}
\begin{figure}[p!]
\begin{center}
\includegraphics[width=8.5cm]{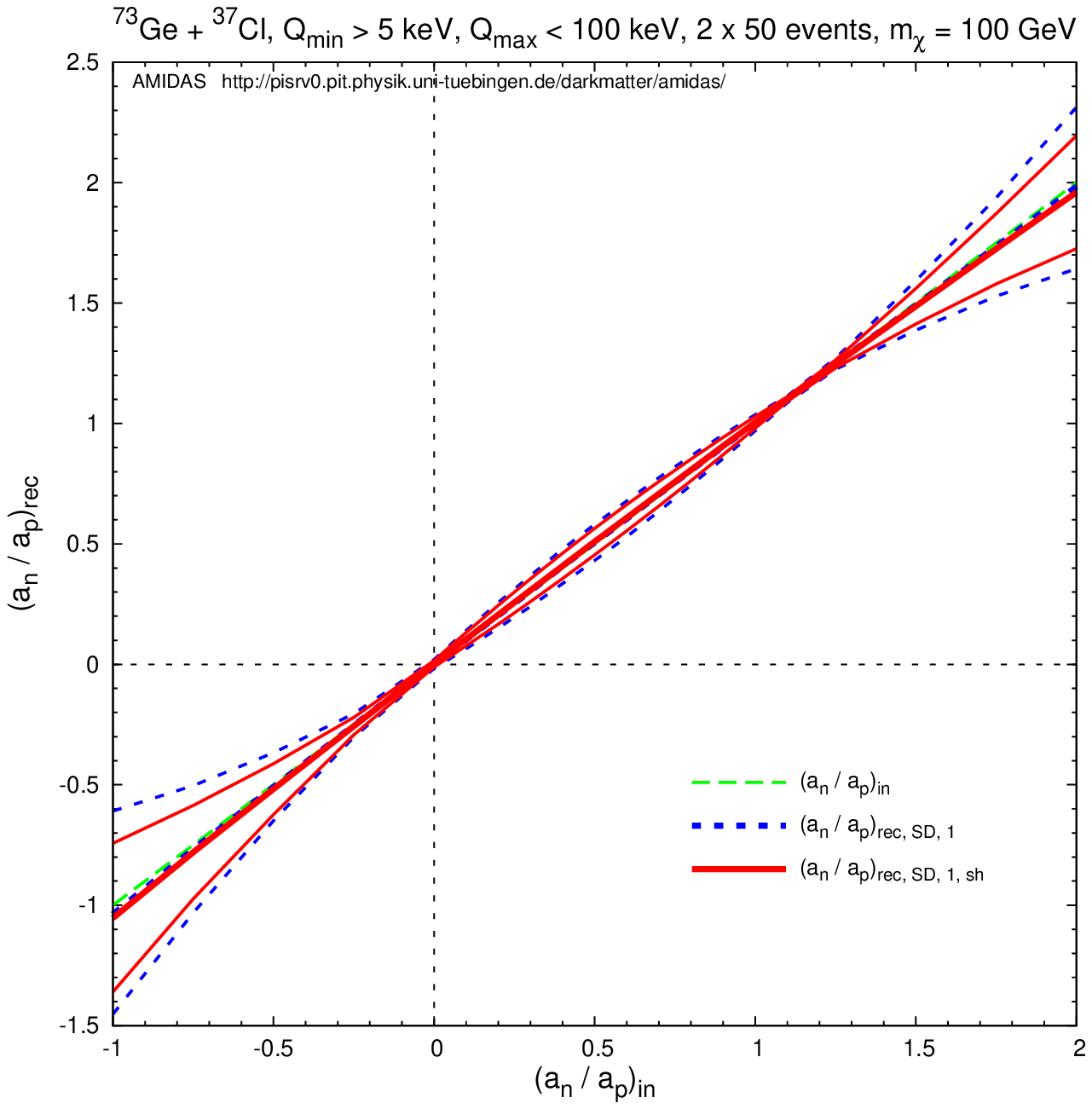}
\includegraphics[width=8.5cm]{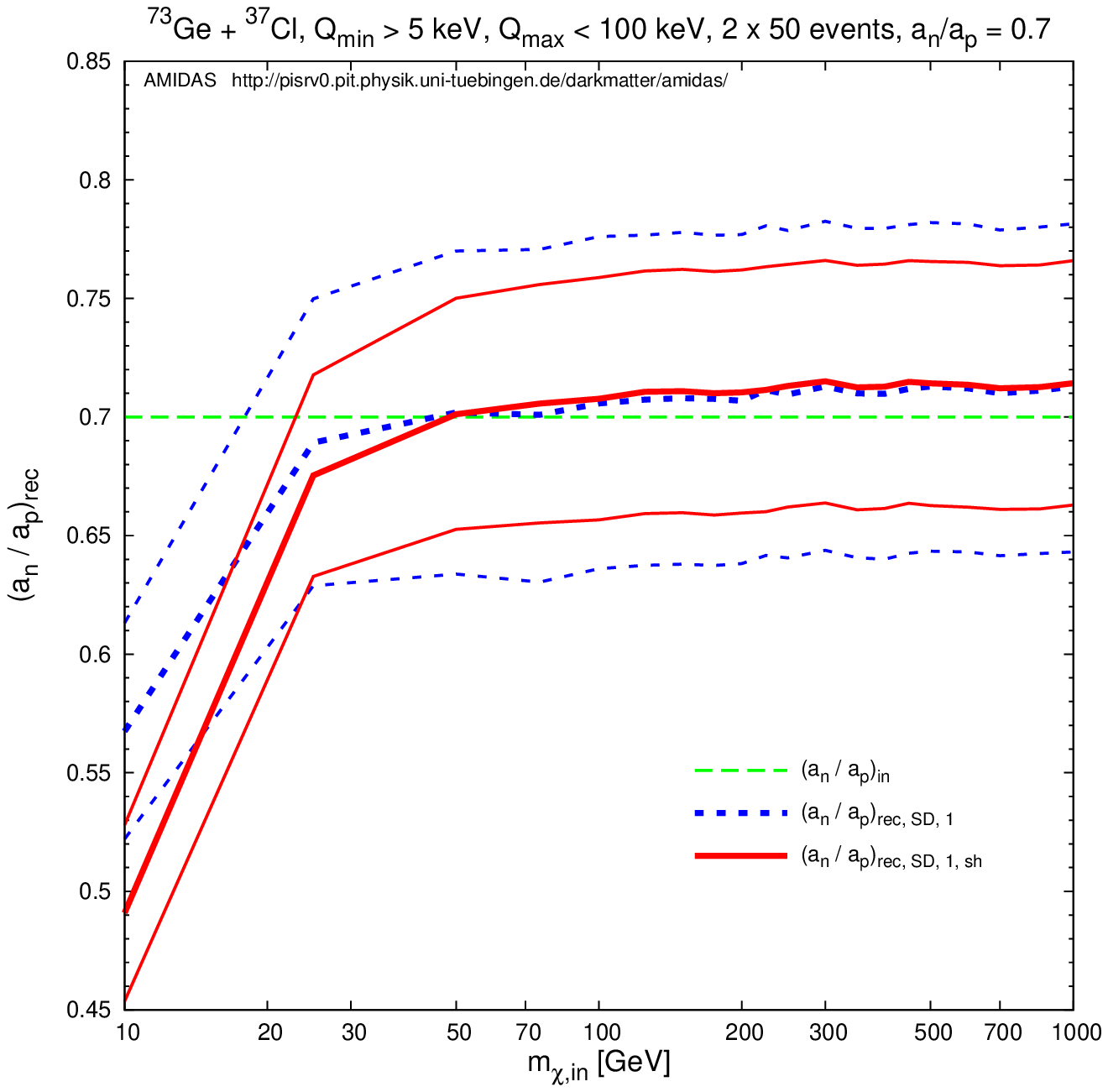}
\vspace{-0.75cm}
\end{center}
\caption{
 Left:
 the combined results of the ``$+$'' and ``$-$'' solutions
 estimated with $r_{(X, Y)}(Q_{{\rm min}, (X, Y)})$ (dashed blue)
 shown in Figs.~\ref{fig:ranapSD_ranap_pm}
 and with $r_{(X, Y)}(Q_{s, 1, (X, Y)})$ (solid red)
 in Figs.~\ref{fig:ranapSD_ranap_sh_pm}.
 Right:
 a comparison of the results
 shown in Figs.~\ref{fig:ranapSD_mchi_rec}
 estimated with $r_{(X, Y)}(Q_{{\rm min}, (X, Y)})$ (dashed blue)
 and with with $r_{(X, Y)}(Q_{s, 1, (X, Y)})$ (solid red).
 Only the results estimated with $n = 1$
 are shown here.
}
\label{fig:ranapSD_ranap_mchi}
\end{figure}

 On the other hand,
 as mentioned above,
 the expression (\ref{eqn:ranapSD}) for estimating
 the ratio between two SD WIMP--nucleon couplings
 is independent of the WIMP mass.
 In Figs.~\ref{fig:ranapSD_mchi_rec},
 I show the reconstructed $\armn / \armp$ ratio
 and the lower and upper bounds of
 their 1$\sigma$ statistical uncertainties
 as functions of the {\em input WIMP mass} $\mchi$
 for a fixed input $\armn / \armp = 0.7$.
 We estimate with $r_{(X, Y)}(Q_{{\rm min}, (X, Y)})$
 and $r_{(X, Y)}(Q_{s, 1, (X, Y)})$
 in the left and right frames,
 respectively.
 It can be seen that,
 firstly,
 except the statistical uncertainty
 estimated with $r_{(X, Y)}(Q_{{\rm min}, (X, Y)})$ and \mbox{$n = -1$}
 (the dashed blue curves
  labeled as $(\armn / \armp)_{\rm rec,~SD,~-1}$ in the left frame),
 for WIMP masses \mbox{$\mchi~\gsim$ 50 GeV},
 the reconstructed $\armn / \armp$ ratio
 as well as the statistical uncertainty
 are (almost) independent of the WIMP mass;
 however,
 if WIMPs are (very) light (\mbox{$\mchi~\lsim~25$ GeV}),
 $\armn / \armp$ will be (strongly) underestimated,
 due to the non--zero threshold energies%
\footnote{
 Remind that,
 as discussed in Ref.~\cite{DMDDfp2}
 for the method for estimating the SI WIMP--nucleon coupling,
 this kind of underestimate
 (or overestimate shown later in this article)
 can be alleviated
 (corrected)
 once we can decrease the threshold energies
 (to be negligible);
 see also Ref.~\cite{DMDDbg-ranap}
 for simulations with {\em negligible}
 experimental threshold energies.
}.
 Secondly,
 the statistical uncertainties on $\armn / \armp$
 estimated with $r_{(X, Y)}(Q_{{\rm min}, (X, Y)})$
 and $r_{(X, Y)}(Q_{s, 1, (X, Y)})$
 are only 10\% or even 7\%
 combined with an \mbox{$\sim$ 1.5\%} systematic deviation.

 As a comparison,
 I show the combinations of the ``$+$'' and ``$-$'' solutions with $n = 1$
 shown in Figs.~\ref{fig:ranapSD_ranap_pm}
 and \ref{fig:ranapSD_ranap_sh_pm}
 together in the left frame of Figs.~\ref{fig:ranapSD_ranap_mchi}.
 In the right frame,
 I compare also the results with $n = 1$
 shown in Figs.~\ref{fig:ranapSD_mchi_rec}.
 The $\sim 30\%$ (from 10\% to 7\%) reduction of
 the statistical uncertainty
 by estimating with $r_{(X, Y)}(Q_{s, 1, (X, Y)})$
 for $\mchi~\gsim~100$ GeV
 can be seen obviously.

\begin{figure}[p!]
\begin{center}
\includegraphics[width=8.5cm]{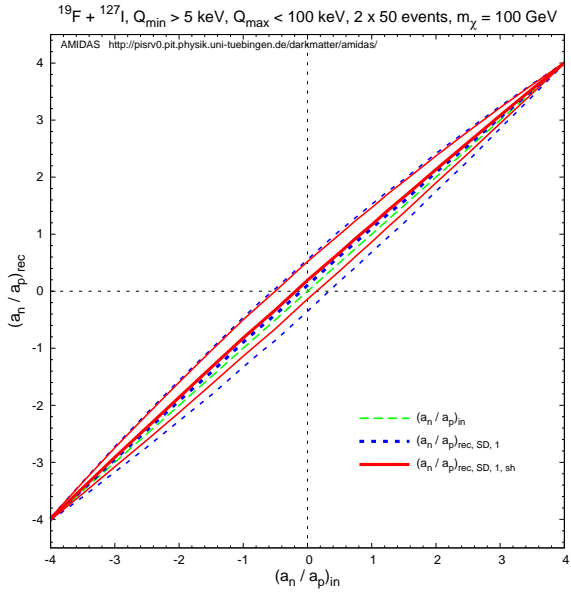}
\includegraphics[width=8.5cm]{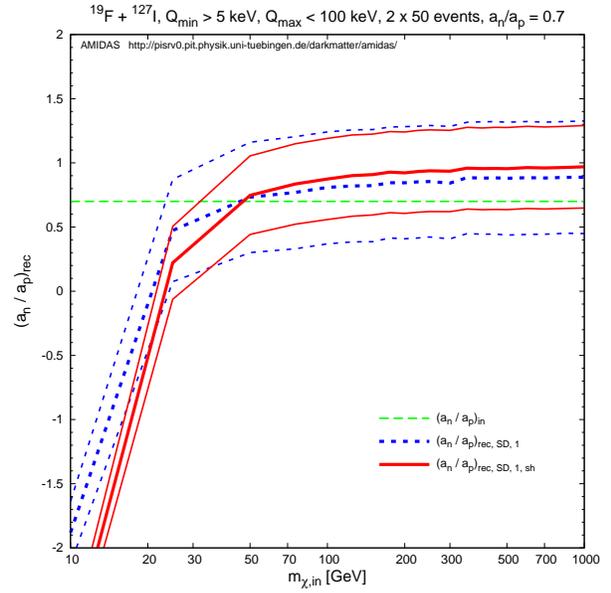}
\vspace{-0.75cm}
\end{center}
\caption{
 As in Figs.~\ref{fig:ranapSD_ranap_mchi},
 except that
 $\rmXA{F}{19}$ and $\rmXA{I}{127}$
 have been chosen as two target nuclei.
 Note that the scales used here
 are different from those
 in Figs.~\ref{fig:ranapSD_ranap_mchi}.
}
\label{fig:ranapSD_ranap_mchi_fi}
\end{figure}

 Furthermore,
 considering the low natural abundances of
 $\rmXA{Ge}{73}$ and $\rmXA{Cl}{37}$
 (see Table 1),
 in Figs.~\ref{fig:ranapSD_ranap_mchi_fi}
 we simulate with another combination of target nuclei:
 $\rmXA{F}{19}$ and $\rmXA{I}{127}$.
 As discussed in the previous subsection
 and shown in Figs.~\ref{fig:ranapSD_ranap_pm},
 \ref{fig:ranapSD_ranap_sh_pm},
 and \ref{fig:ranapSD_ranap_mchi},
 the inner solutions of $(\armn / \armp)_{\pm, n}^{\rm SD}$
 have a much smaller statistical uncertainties
 and the range of these inner solutions
 depends on the $- \Srmp / \Srmn$ values of our target nuclei.
 Hence,
 one benefit of using the combination of
 $\rmXA{F}{19}$ and $\rmXA{I}{127}$ is that
 one can estimate $(\armn / \armp)_{\pm, n}^{\rm SD}$
 in a much wilder range of interest:
 $|\armn / \armp| \le 4$.
 Consequently,
 for the practical use of analyzing real data,
 one has therefore {\em not} to worry about
 making the choice from the ``$+$'' and ``$-$'' estimates,
 which is discussed at the end of the previous subsection;
 since $\Srmn_{\rmXA{F}{19}}$ and $\Srmn_{\rmXA{I}{127}}$
 have different signs,
 we can just take the ``$-$'' solution in Eq.~(\ref{eqn:ranapSD}).

 However,
 Figs.~\ref{fig:ranapSD_ranap_mchi_fi} show us also
 some drawbacks of the use of
 the $\rmXA{F}{19}$ $+$ $\rmXA{I}{127}$ combination.
 For WIMP masses $\mchi~\gsim~50$ GeV,
 $(\armn / \armp)_{\pm, n}^{\rm SD}$ estimated
 with $r_{(X, Y)}(Q_{{\rm min}, (X, Y)})$ (dashed blue)
 are \mbox{$\sim$ 15\% - 30\%} {\em overestimated};
 whereas those estimated
 with $r_{(X, Y)}(Q_{s, 1, (X, Y)})$ (solid red) are even worse:
 $\sim 36\%$ for $\mchi \sim 1$ TeV.
 Moreover,
 the statistical uncertainties shown here become also
 much larger (of a factor of $\sim 3 - 5$)
 than those shown in Figs.~\ref{fig:ranapSD_ranap_mchi}.
 This enlargement of the statistical uncertainties
 is mainly caused by the larger value of
 the prefactor of $\sigma((\armn / \armp)_{\pm, n}^{\rm SD})$
 in Eq.~(\ref{eqn:sigma_ranapSD}).
 According to Table 1,
 the values of $\vBig{\SpY \SnX - \SpX \SnY}$
 are $\sim$ 0.023 for the \mbox{Ge $+$ Cl} combination,
 but $\sim$ 0.067 for F $+$ I.
 Meanwhile,
 as shown in both Figs.~\ref{fig:ranapSD_ranap_mchi}
 and \ref{fig:ranapSD_ranap_mchi_fi},
 the statistical uncertainties are at the largest
 in the middle of two coincidence points
 and reduce as the input $\armn / \armp$
 approaches to one of these two points.
 Since we set the input $\armn / \armp = 0.7$
 for simulations with different WIMP masses,
 comparing to the relative difference between 0.7
 and the middle point of 1.16 and $-$0.08,
 i.e., 0.54,
 the relative difference between 0.7
 and the middle point of 4.05 and $-$4.12,
 i.e., $-$0.035,
 is slightly {\em smaller}.
 This causes also a {\em larger} statistical uncertainty
 for the use of the F $+$ I combination.
 In contrast,
 for light WIMP masses (\mbox{$\mchi~\lsim~50$ GeV}),
 the two estimates with $\rmXA{F}{19}$ and $\rmXA{I}{127}$
 shown in the right frame of Figs.~\ref{fig:ranapSD_ranap_mchi_fi}
 are much more underestimated than results
 shown in the right frame of Figs.~\ref{fig:ranapSD_ranap_mchi}.
\section{Combination of the SI and SD cross sections}
 In this section
 I consider the general combination
 of the SI and SD WIMP--nucleus cross sections.
\subsection{General expression}
 At first,
 by combining Eqs.~(\ref{eqn:sigma0SI}), (\ref{eqn:sigma0SD}),
 and (\ref{eqn:sigmap/nSD}),
 we can find
\beq
   \frac{\sigmaSD}{\sigmaSI}
 = \afrac{32}{\pi} G_F^2 \~ \mrp^2 \Afrac{J + 1}{J}
   \bfrac{\Srmp + \Srmn (\armn / \armp)}{A}^2 \frac{|\armp|^2}{\sigmapSI}
 = \calCp \afrac{\sigmapSD}{\sigmapSI}
\~,
\label{eqn:rsigmaSDSI}
\eeq
 where I have defined
\beq
        \calCp
 \equiv \frac{4}{3} \afrac{J + 1}{J}
        \bfrac{\Srmp + \Srmn (\armn/\armp)}{A}^2
\~.
\label{eqn:Cp}
\eeq
 For the general combination of
 the SI and SD WIMP--nucleus cross sections,
 the expression (\ref{eqn:dRdQ})
 for the differential event rate
 should be modified to
\beqn
        \adRdQ_{\rm expt}
 \=     \calE \afrac{\rho_0}{2 \mchi \mrN^2}
        \bbigg{\sigmaSI \FSIQ + \sigmaSD \FSDQ}
        \int_{\vmin}^{\vmax}
        \bfrac{f_1(v)}{v} dv
        \non\\
 \=     \calE
        A^2 \! \afrac{\rho_0 \sigmapSI}{2 \mchi \mrp^2} \!\!
        \bbrac{\FSIQ + \afrac{\sigmapSD}{\sigmapSI} \calCp \FSDQ}
        \int_{\vmin}^{\vmax}
        \bfrac{f_1(v)}{v} dv
\~,
\label{eqn:dRdQ_SISD}
\eeqn
 where I have used Eq.~(\ref{eqn:sigma0SI}) again.
 Then one can find straightforwardly that
 the integral above can be estimated by Eq.~(\ref{eqn:moments})
 with the following replacement:
\beq
        \FQmin
 \to     F'^2(\Qmin)
 \equiv  F_{\rm SI}^2(\Qmin)
       + \abrac{\sigmapSD / \sigmapSI} \calCp F_{\rm SD}^2(\Qmin)
\~.
\eeq
 Hence,
 for this general case,
 Eq.~(\ref{eqn:rho_sigma}) becomes to
\beq
    \rho_0
    \bbrac{A^2 \afrac{\mrN}{\mrp}^2 \sigmapSI}
 = \afrac{1}{\calE}
   \mchi \mrN \sfrac{\mN}{2}
   \bbrac{\frac{2 \Qmin^{1/2} r(\Qmin)}{F'^2(\Qmin)} + I_0}
\~,
\label{eqn:rho_sigma_SISD}
\eeq
 where
\beq
   I_n(\Qmin, \Qmax)
 = \sum_{a = 1}^{N_{\rm tot}} \frac{Q_a^{(n-1)/2}}{F'^2(Q_a)}
\~.
\label{eqn:In_sum_SISD}
\eeq

 Now by combining two targets $X$ and $Y$
 and using the definition (\ref{eqn:alpha}) of $\alpha$,
 the relation (\ref{eqn:ralphaXY}) between $\alpha_X / \alpha_Y$
 with $n = -1$,
 as well as the expression (\ref{eqn:RnX_min}) for $\calR_{n, (X, Y)}$,
 one can obtain that%
\footnote{
 This equation can be obtained
 by simply assuming that
 the integral over $f_1(v)$
 on the right--hand side of Eq.~(\ref{eqn:dRdQ_SISD})
 estimated in two experiments
 (approximately) agree
 and can thus be cancelled by each other.
 This assumption can practically always hold,
 even though the experimental minimal and maximal cut--off energies
 in these two experiments
 should be matched by requiring \cite{DMDDmchi} that
\(
   \alpha_X \sqrt{Q_{{\rm (min, max)}, X}}
 = \alpha_Y \sqrt{Q_{{\rm (min, max)}, Y}}
\),
 since,
 as the expressions (\ref{eqn:rsigmaSDpSI}) and (\ref{eqn:rsigmaSDnSI}) show,
 only the estimated values of $r_{(X, Y)}(Q_{{\rm min}, (X, Y)})$
 are important for the data analysis.
 Note that,
 however,
 once one applies similarly this simple cancellation
 for the case of a dominant SD WIMP cross section
 discussed in the previous section,
 only the expression (\ref{eqn:ranapSD})
 for $(\armn / \armp)_{\pm, n}^{\rm SD}$ with $n = -1$,
 namely with $\calR_{J, -1, (X, Y)}$ given in Eq.~(\ref{eqn:RJmaX}),
 can be obtained.
 This is because that,
 by using this cancellation,
 $\alpha \propto \sqrt{\mN} / \mrN$ and
 \mbox{$2 \Qmin^{1 / 2} r(\Qmin) / \FQmin + I_0$}
 on the right--hand side of Eq.~(\ref{eqn:rho_sigma})
 will be eliminated {\em before}
 one obtains this equation.
 Then one {\em cannot} use the relation (\ref{eqn:ralphaXY})
 to convert $\alpha_X / \alpha_Y$
 to $\calR_{n, Y} / \calR_{n, X}$
 and therefore to obtain the expression (\ref{eqn:ranapSD})
 with different values of $n$;
 except with $n = -1$,
 since
 \mbox{$2 \QminX^{1 /2} r_X(\QminX) / \FQminX + \IzX$}
 appears in the {\em numerator} of $\calR_{-1, X}$
 (see Eq.~(\ref{eqn:RnX_min}),
  not in the denominator as for the cases with $n = 1,~2,~\cdots$)
 and can thus be cancelled out anyway.
}
\beq
   \frac{\FSIQminX + \abrac{\sigmapSD / \sigmapSI} \calCpX \FSDQminX}
        {\FSIQminY + \abrac{\sigmapSD / \sigmapSI} \calCpY \FSDQminY}
 = \frac{\calR_{m, X}}{\calR_{m, Y}}
\~,
\label{eqn:rFSIQXY'}
\eeq
 where I have assumed $m_{(X, Y)} \propto A_{(X, Y)}$ and defined
\beq
        \calR_{m, X}
 \equiv \frac{r_X(\QminX)}{\calE_X \mX^2}
\~.
\label{eqn:RmX}
\eeq
 From Eq.~(\ref{eqn:rFSIQXY'}),
 the ratio of the SD WIMP--proton cross section
 to the SI one can be solved analytically as
 \cite{DMDDidentification-DMDE2009}
\beq
   \frac{\sigmapSD}{\sigmapSI}
 = \frac{\FSIQminY (\calR_{m, X}/\calR_{m, Y}) - \FSIQminX}
        {\calCpX \FSDQminX - \calCpY \FSDQminY (\calR_{m, X} / \calR_{m, Y})}
\~,
\label{eqn:rsigmaSDpSI}
\eeq
 where ${\cal C}_{{\rm p}, (X, Y)}$ have been defined in Eq.~(\ref{eqn:Cp}).
 Similarly,
 the ratio of the SD WIMP--neutron cross section
 to the SI one can be given analogously as
 \cite{DMDDidentification-DMDE2009}%
\footnote{
 Here I assumed that $\sigmanSI \simeq \sigmapSI$ by Eq.~(\ref{eqn:fp/n}).
}
\beq
   \frac{\sigmanSD}{\sigmapSI}
 = \frac{\FSIQminY (\calR_{m, X}/\calR_{m, Y}) - \FSIQminX}
        {\calCnX \FSDQminX - \calCnY \FSDQminY (\calR_{m, X} / \calR_{m, Y})}
\~,
\label{eqn:rsigmaSDnSI}
\eeq
 with the definition
\beq
        \calCn
 \equiv \frac{4}{3} \Afrac{J + 1}{J}
        \bfrac{\Srmp (\armp/\armn) + \Srmn}{A}^2
\~.
\label{eqn:Cn}
\eeq
 The emphasize here is that
 one can use expressions
 (\ref{eqn:rsigmaSDpSI}) and (\ref{eqn:rsigmaSDnSI})
 to estimate $\sigma_{\chi ({\rm p, n})}^{\rm SD} / \sigmapSI$
 {\em without} a prior knowledge of the WIMP mass $\mchi$.
 Moreover,
 since ${\cal C}_{{\rm (p, n)}, (X, Y)}$
 depend only on the nature of the detector materials,
 $\sigma_{\chi ({\rm p, n})}^{\rm SD} / \sigmapSI$
 are practically only functions of $\calR_{m, (X, Y)}$,
 i.e., the counting rates
 at the experimental minimall cut--off energies,
 which can be estimated by using events
 in the lowest available energy ranges.
\subsection{Using \boldmath$(\armn / \armp)_{\pm, n}^{\rm SD}$
            in Eq.~(\ref{eqn:ranapSD})}
 Since $\calCp$ and $\calCn$
 defined in Eqs.~(\ref{eqn:Cp}) and (\ref{eqn:Cn})
 are functions of $\armn / \armp$,
 once the $\armn / \armp$ ratio has been estimated
 (from e.g., some other direct detection experiments
  by Eq.~(\ref{eqn:ranapSD})
  under the assumption of a dominant SD WIMP--nucleus interaction),
 $\sigmapSD / \sigmapSI$ can then be estimated
 by Eq.~(\ref{eqn:rsigmaSDpSI})
 with the following statistical uncertainty%
\footnote{
 Hereafter I consider only the case with protons.
 But all formulae given in this section
 can be applied straightforwardly to the case with neutrons
 by replacing p $\to$ n and $\calCp \to \calCn$.
}:
\beqn
        \sigma\afrac{\sigmapSD}{\sigmapSI}
 \=     \cleft{    \bbrac{\pp{(\armn / \armp)} \afrac{\sigmapSD}{\sigmapSI}}^2
                   \sigma^2\abrac{\afrac{\armn}{\armp}_{\pm, n}^{\rm SD}}}
        \non\\
 \conti ~~~~~~~~~~~~ 
        \cright{+ \sum_{i = X, Y}
                  \bbrac{  \frac{1}{\calE_i m_i^2}~
                           \pp{\calR_{m, i}} \afrac{\sigmapSD}{\sigmapSI}}^2
                  \sigma^2(r_i(Q_{{\rm min}, i})) }^{1/2}
\~,
\label{eqn:sigma_rsigmaSDpSI_ranapSD}
\eeqn
 where
\beq
    \pp{(\armn / \armp)} \afrac{\sigmapSD}{\sigmapSI}
 =  \pp{\calCpX} \afrac{\sigmapSD}{\sigmapSI} \cdot
    \Pp{\calCpX}{(\armn / \armp)}
  + \pp{\calCpY} \afrac{\sigmapSD}{\sigmapSI} \cdot
    \Pp{\calCpY}{(\armn / \armp)}
\~.
\label{eqn:drsigmaSDpSI_dranapSISD}
\eeq
 Explicit derivatives of $\sigmapSD / \sigmapSI$
 with respect to ${\cal C}_{{\rm p}, (X, Y)}$ and $\calR_{m, (X, Y)}$
 will be given in the appendix.
 Note that
 Eq.~(\ref{eqn:ranapSD}) can be used
 only when the SD WIMP--nucleus interaction
 really dominates over the SI one.
 We will see later that,
 if the SD interaction {\em does not} dominate,
 the $\armn / \armp$ ratio
 should not be estimated by Eq.~(\ref{eqn:ranapSD}) any more.
\subsection{Solving \boldmath$\armn / \armp$ with a third nucleus}
 Nevertheless,
 for the general combination of
 the SI and SD WIMP--nucleus cross sections,
 the $\armn / \armp$ ratio can in fact be solved analytically
 by introducing a {\em third nucleus}
 with {\em only} an SI sensitivity:
\beq
   \Srmp_Z
 = \Srmn_Z
 = 0
\~,
\label{eqn:Sp/nZ}
\eeq
 i.e.,
\beq
   {\cal C}_{{\rm p}, Z}
 = 0
\~.
\label{eqn:CpZ}
\eeq
 Then,
 according to Eq.~(\ref{eqn:rsigmaSDpSI}),
 we have
\beqN
   \frac{\FSIQminZ (\calR_{m, X} / \calR_{m, Z}) - \FSIQminX}{\calCpX \FSDQminX}
 = \frac{\FSIQminZ (\calR_{m, Y} / \calR_{m, Z}) - \FSIQminY}{\calCpY \FSDQminY}
\~.
\eeqN
 Using $\calCp$ defined in Eq.~(\ref{eqn:Cp}),
 the $\armn / \armp$ ratio can be solved analytically as
 \cite{DMDDidentification-DMDE2009}
\beqn
    \afrac{\armn}{\armp}_{\pm}^{\rm SI + SD}
 \= \frac{-\abrac{\cpX \snpX - \cpY \snpY}
          \pm \sqrt{\cpX \cpY} \vbrac{\snpX - \snpY}}
         {\cpX \snpX^2 - \cpY \snpY^2}
    \non\\
 \= \cleft{\renewcommand{\arraystretch}{0.5}
           \begin{array}{l l l}
            \\
            \D -\frac{\sqrt{\cpX} \mp \sqrt{\cpY}}{\sqrt{\cpX} \snpX \mp \sqrt{\cpY} \snpY}\~, &
            ~~~~~~~~ & ({\rm for}~\snpX > \snpY), \\~\\~\\ 
            \D -\frac{\sqrt{\cpX} \pm \sqrt{\cpY}}{\sqrt{\cpX} \snpX \pm \sqrt{\cpY} \snpY}\~, &
                     & ({\rm for}~\snpX < \snpY). \\~\\
           \end{array}}
\label{eqn:ranapSISD}
\eeqn
 Here I have defined
\cheqna
\beq
        \cpX
 \equiv \frac{4}{3} \Afrac{J_X + 1}{J_X} \bfrac{\SpX}{A_X}^2
        \bbrac{  \FSIQminZ \afrac{\calR_{m, Y}}{\calR_{m, Z}} \!
               - \FSIQminY} \!
        \FSDQminX
\~,
\label{eqn:cpX}
\eeq
\cheqnb
\beq
        \cpY
 \equiv \frac{4}{3} \Afrac{J_Y + 1}{J_Y} \bfrac{\SpY}{A_Y}^2
        \bbrac{  \FSIQminZ \afrac{\calR_{m, X}}{\calR_{m, Z}} \!
               - \FSIQminX} \!
        \FSDQminY
\~,
\label{eqn:cpY}
\eeq
\cheqn
 and
\beq
        \snpX
 \equiv \frac{\SnX}{\SpX}
\~.
\label{eqn:snpX}
\eeq
 Note that,
 firstly,
 $(\armn / \armp)_{\pm}^{\rm SI + SD}$ and $c_{{\rm p}, (X, Y)}$
 given in Eqs.~(\ref{eqn:ranapSISD}), (\ref{eqn:cpX}), and (\ref{eqn:cpY})
 are functions of only
 $r_{(X, Y, Z)}(Q_{{\rm min}, (X, Y, Z)})$,
 which can be estimated with events
 in the lowest energy ranges.
 Secondly,
 while the decision of the inner solution of
 $(\armn / \armp)_{\pm, n}^{\rm SD}$
 depends on the signs of $\SnX$ and $\SnY$,
 the decision with $(\armn / \armp)_{\pm}^{\rm SI + SD}$
 depends {\em not only} on the signs of
 \mbox{$\snpX = \SnX / \SpX$} and \mbox{$\snpY = \SnY / \SpY$},
 {\em but also} on the {\em order} of the two targets.
 For the Ge + Cl combination,
 since \mbox{$ s_{{\rm n/p}, \rmXA{Ge}{73}} = 12.6
             > s_{{\rm n/p}, \rmXA{Cl}{37}} = -0.86$},
 one should use the {\em upper} expression
 in the second line of Eq.~(\ref{eqn:ranapSISD}),
 and since $s_{{\rm n/p}, \rmXA{Ge}{73}}$
 and $s_{{\rm n/p}, \rmXA{Cl}{37}}$
 have the opposite signs,
 the ``$-$ (minus)'' solution of this expression
 (or the ``$+$ (plus)'' solution of the expression in the first line)
 is the inner solution.
 In contrast,
 since \mbox{$ s_{{\rm n/p}, \rmXA{F}{19}}  = -0.247
             < s_{{\rm n/p}, \rmXA{I}{127}} = 0.243$}
 and since $s_{{\rm n/p}, \rmXA{F}{19}}$
 and $s_{{\rm n/p}, \rmXA{I}{127}}$
 have the opposite signs,
 the ``$-$ (minus)'' solution of the {\em lower} expression
 in the second line of Eq.~(\ref{eqn:ranapSISD})
 (or the ``$-$ (minus)'' solution of the expression in the first line)
 is then the inner solution
 for the F + I combination.

 Finally,
 from the expression (\ref{eqn:ranapSISD}),
 the statistical uncertainty
 on $(\armn / \armp)_{\pm}^{\rm SI + SD}$
 can be given by
\beqn
        \sigma\abrac{\afrac{\armn}{\armp}_{\pm}^{\rm SI + SD}}
 \=     \cleft{   \sum_{i = X, Y, Z}
                  \bleft{   \pp{\cpX}\afrac{\armn}{\armp}_{\pm}^{\rm SI + SD} \cdot
                            \Pp{\cpX}{r_i(Q_{{\rm min}, i})}}}
        \non\\
 \conti ~~~~~~~~~~~~~~~~ 
        \cright{  \bright{+ \pp{\cpY}\afrac{\armn}{\armp}_{\pm}^{\rm SI + SD} \cdot
                            \Pp{\cpY}{r_i(Q_{{\rm min}, i})}}^2
                  \sigma^2(r_i(Q_{{\rm min}, i}))}^{1/2}
\!\!.
\label{eqn:sigma_ranapSISD}
\eeqn
 And the statistical uncertainty on
 the ratio between two WIMP--proton cross sections
 in Eq.~(\ref{eqn:rsigmaSDpSI}) can be expressed as
 (c.f., Eq.~(\ref{eqn:sigma_rsigmaSDpSI_ranapSD}))
\beqn
        \sigma\afrac{\sigmapSD}{\sigmapSI}
 \=     \cBiggl{  \sum_{i = X, Y, Z}
        \cleft{   \bbrac{\pp{(\armn / \armp)} \afrac{\sigmapSD}{\sigmapSI}}}}
                  \bbrac{\pp{r_i(Q_{{\rm min}, i})}
                         \afrac{\armn}{\armp}_{\pm}^{\rm SI + SD}}
        \non\\
 \conti ~~~~~~~~~~~~~~~~~~~~~~~~ 
        \cBiggr{
        \cright{+ \frac{1}{\calE_i m_i^2}
                  \bbrac{\pp{\calR_{m, i}} \afrac{\sigmapSD}{\sigmapSI}}}^2
                  \sigma^2(r_i(Q_{{\rm min}, i}))}^{1/2}
\~,
\label{eqn:sigma_rsigmaSDpSI_ranapSISD}
\eeqn
 with $\p \abrac{\sigmapSD / \sigmapSI} / \p (\armn / \armp)$
 given in Eq.~(\ref{eqn:drsigmaSDpSI_dranapSISD})
 and
\beq
    \pp{r_i(Q_{{\rm min}, i})} \afrac{\armn}{\armp}_{\pm}^{\rm SI + SD} \!\!\!\!
 =  \pp{\cpX} \afrac{\armn}{\armp}_{\pm}^{\rm SI + SD} \!\!\!\! \cdot
    \Pp{\cpX}{r_i(Q_{{\rm min}, i})}
  + \pp{\cpY} \afrac{\armn}{\armp}_{\pm}^{\rm SI + SD} \!\!\!\! \cdot
    \Pp{\cpY}{r_i(Q_{{\rm min}, i})}
\~,
\label{eqn:dranapSISD_drminX}
\eeq
 for $i = X,~Y,~Z$.
 Explicit derivatives of $(\armn / \armp)_{\pm}^{\rm SI + SD}$
 and $c_{{\rm p}, (X, Y)}$
 will be given in the appendix.

\begin{figure}[t!]
\begin{center}
\includegraphics[width=8.5cm]{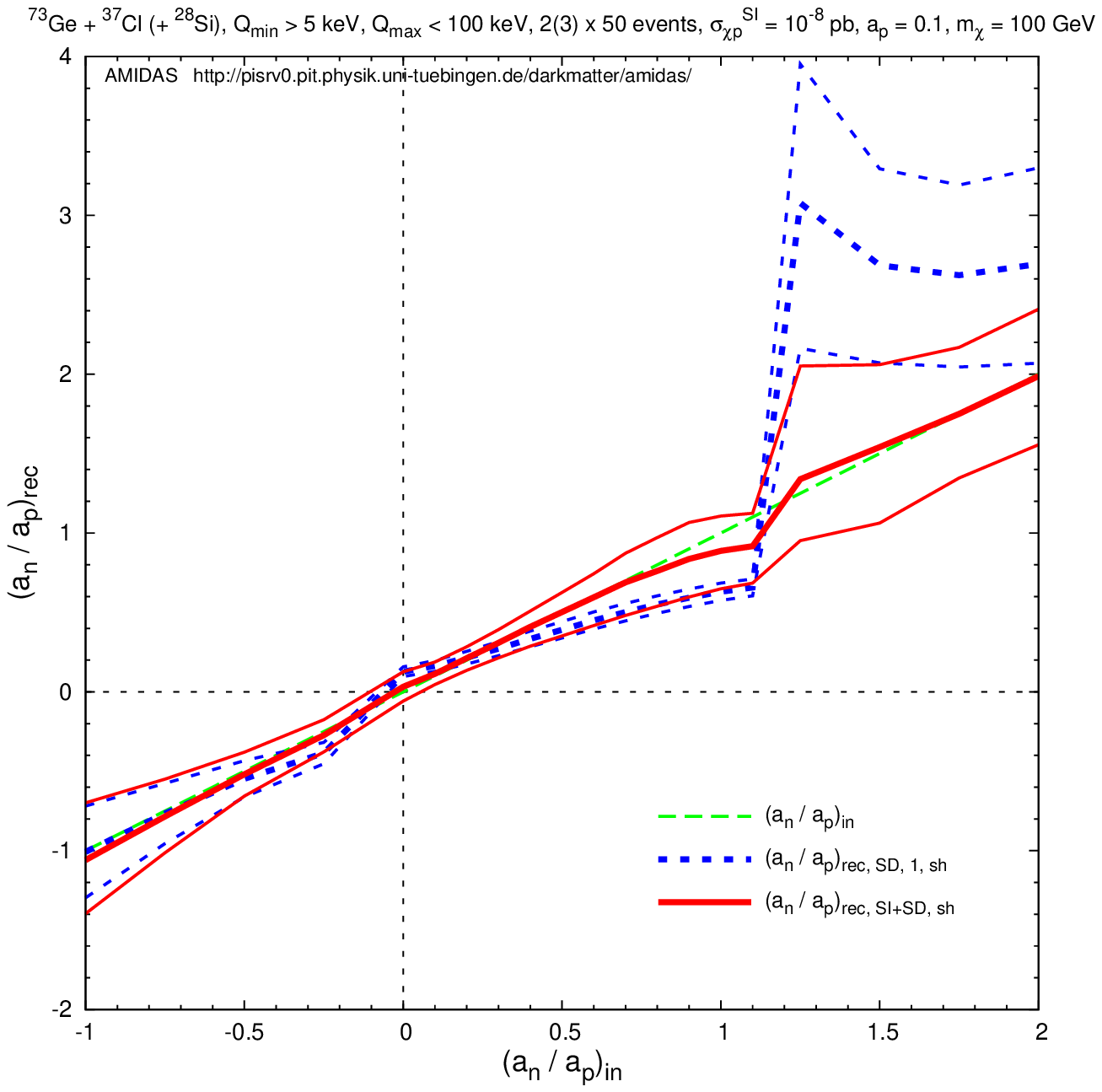}
\includegraphics[width=8.5cm]{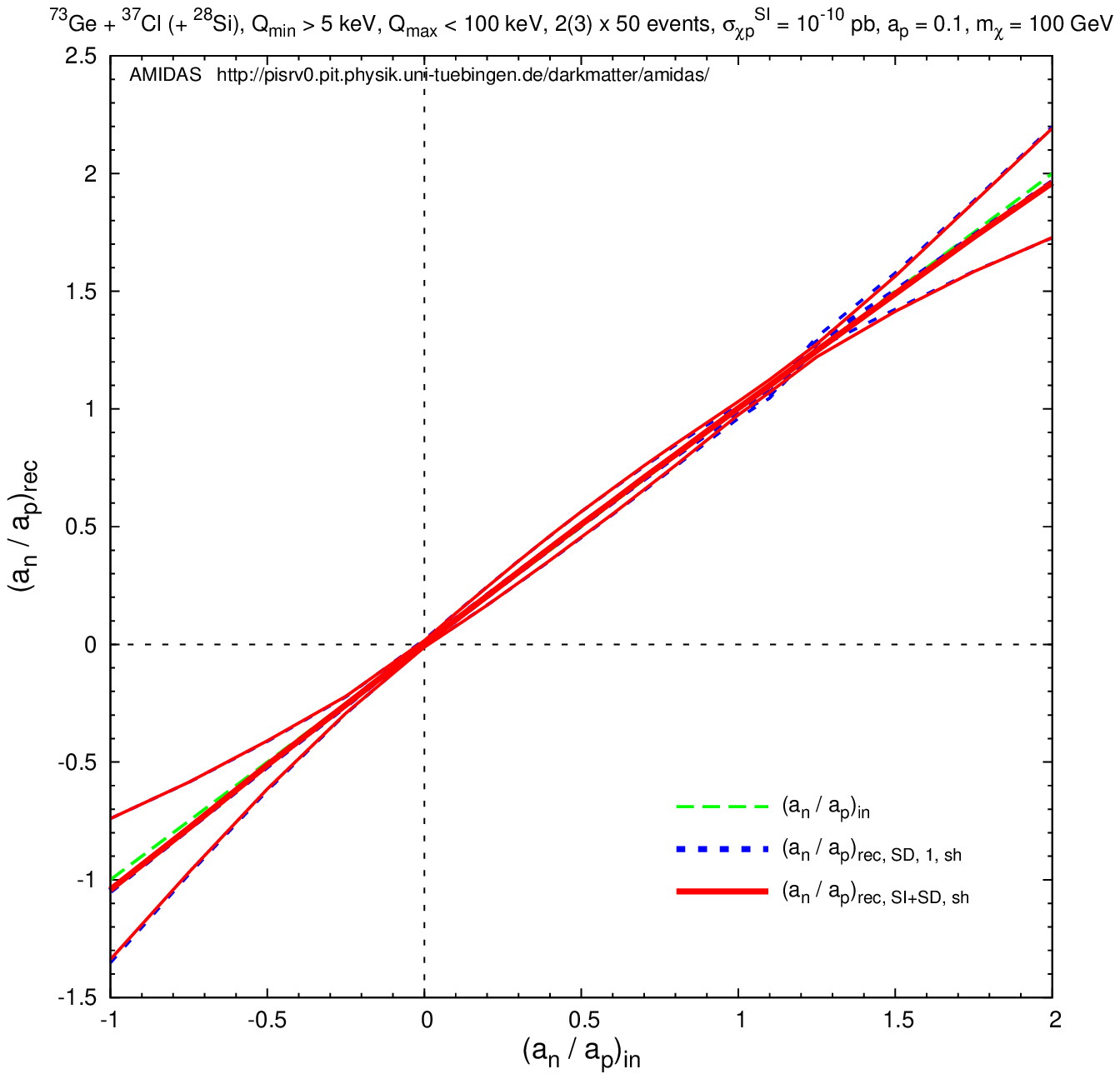}
\vspace{-0.75cm}
\end{center}
\caption{
 The reconstructed $\armn / \armp$ ratios
 estimated by Eqs.~(\ref{eqn:ranapSD}) (dashed blue)
 and (\ref{eqn:ranapSISD}) (solid red)
 and the lower and upper bounds of
 their 1$\sigma$ statistical uncertainties
 estimated by Eqs.~(\ref{eqn:sigma_ranapSD})
 and (\ref{eqn:sigma_ranapSISD})
 as functions of the input $\armn / \armp$ ratio.
 Besides $\rmXA{Ge}{73}$ and $\rmXA{Cl}{37}$,
 $\rmXA{Si}{28}$ has been chosen as the third target
 for estimating $c_{{\rm p}, (X, Y)}$ by
 Eqs.~(\ref{eqn:cpX}) and (\ref{eqn:cpY}).
 For the SI cross section
 the nuclear form factor given in Eq.~(\ref{eqn:FQ_WS})
 has been used.
 The SI WIMP--proton cross section
 has been set as $10^{-8}$ pb (left) and $10^{-10}$ pb (right),
 respectively,
 whereas the SD WIMP--proton coupling $\armp$
 has been set as 0.1.
 The other parameters are as
 in Figs.~\ref{fig:ranapSD_ranap_mchi}.
 Note here that
 the scales of the $(\armn / \armp)_{\rm rec}$--axes
 in two frames are different.
 See the text for further details.
}
\label{fig:ranapSISD_ranap}
\end{figure}

 In Figs.~\ref{fig:ranapSISD_ranap}
 I show the reconstructed $\armn / \armp$ ratios
 estimated by Eqs.~(\ref{eqn:ranapSD}) (dashed blue)
 and (\ref{eqn:ranapSISD}) (solid red)
 and the lower and upper bounds of
 their 1$\sigma$ statistical uncertainties
 estimated by Eqs.~(\ref{eqn:sigma_ranapSD})
 and (\ref{eqn:sigma_ranapSISD})
 as functions of the input $\armn / \armp$ ratio%
\footnote{
 Note that
 all results shown in this subsection
 are only reconstructed with
 \mbox{$r_{(X, Y, Z)}(Q_{s, 1, (X, Y, Z)}) = r_{(X, Y, Z), 1}$}.
}.
 For the SI cross section
 the nuclear form factor given in Eq.~(\ref{eqn:FQ_WS})
 has been used.
 The SI WIMP--proton cross section
 has been set as $10^{-8}$ pb (left) and $10^{-10}$ pb (right),
 respectively,
 whereas the SD WIMP--proton coupling $\armp$
 has been set as 0.1.%
\footnote{
 Remind that
 the current exclusion limit
 on the SI WIMP--nucleon cross section
 is $\lsim~5 \times 10^{-8}$ pb
 for WIMP masses of \mbox{$\sim$ 30 -- 100 GeV}
 from the XENON10 \cite{Angle07},
 CDMS-II \cite{Ahmed09b},
 XENON100 \cite{Aprile10b},
 and EDELWEISS-II \cite{Kozlov10} experiments,
 whereas the limits on the SD WIMP couplings
 on protons and on neutrons are
 $|\armp|~\lsim~0.4$ and $|\armn|~\lsim~0.2$
 (for a WIMP mass of 50 GeV$/c^2$)
 \cite{Felizardo10},
 respectively.
 On the other hand,
 the theoretically predicted values for $\armp$
 is $|\armp|~\lsim~0.1$ \cite{Cotta09}.
}
 Besides $\rmXA{Ge}{73}$ and $\rmXA{Cl}{37}$,
 $\rmXA{Si}{28}$ has been chosen as the third target
 for estimating $c_{{\rm p}, (X, Y)}$ by
 Eqs.~(\ref{eqn:cpX}) and (\ref{eqn:cpY}).

 In the left frame,
 it can be seen obviously that
 $\armn / \armp$ estimated by Eq.~(\ref{eqn:ranapSD}) (dashed blue)
 under the assumption of a dominant SD WIMP--nucleus interaction
 has two discontinuities around
 $(\armn / \armp)_{\rm in} = 1.16$ and $-0.08$
 and the reconstructed $\armn / \armp$ ratio
 is systematically over--/underestimated.
 In contrast,
 $\armn / \armp$ determined by Eq.~(\ref{eqn:ranapSISD}) (solid red)
 shows a more smooth estimate,
 although the reconstructed ratio is a bit {\em underestimated}
 with a relatively larger statistical uncertainty
 for input $\armn / \armp$ ratios around 1.16.
 However,
 once we set the input SI WIMP--proton cross section
 two orders of magnitude lower
 and thus the SD WIMP--nucleus cross section really dominates
 (the right frame),
 the $\armn / \armp$ ratios estimated by two methods
 show a clear compatibility.

\begin{figure}[p!]
\begin{center}
\includegraphics[width=8.5cm]{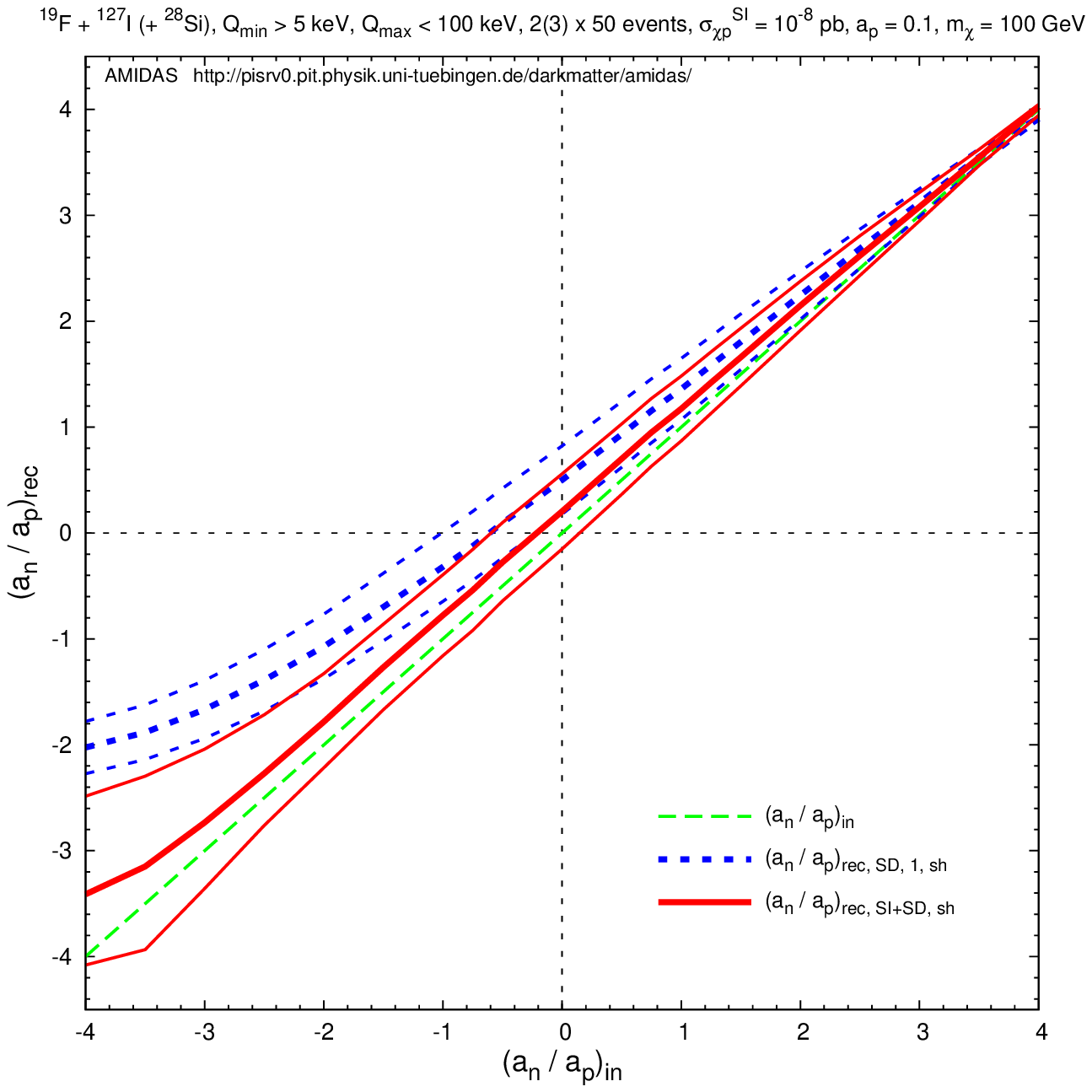}
\includegraphics[width=8.5cm]{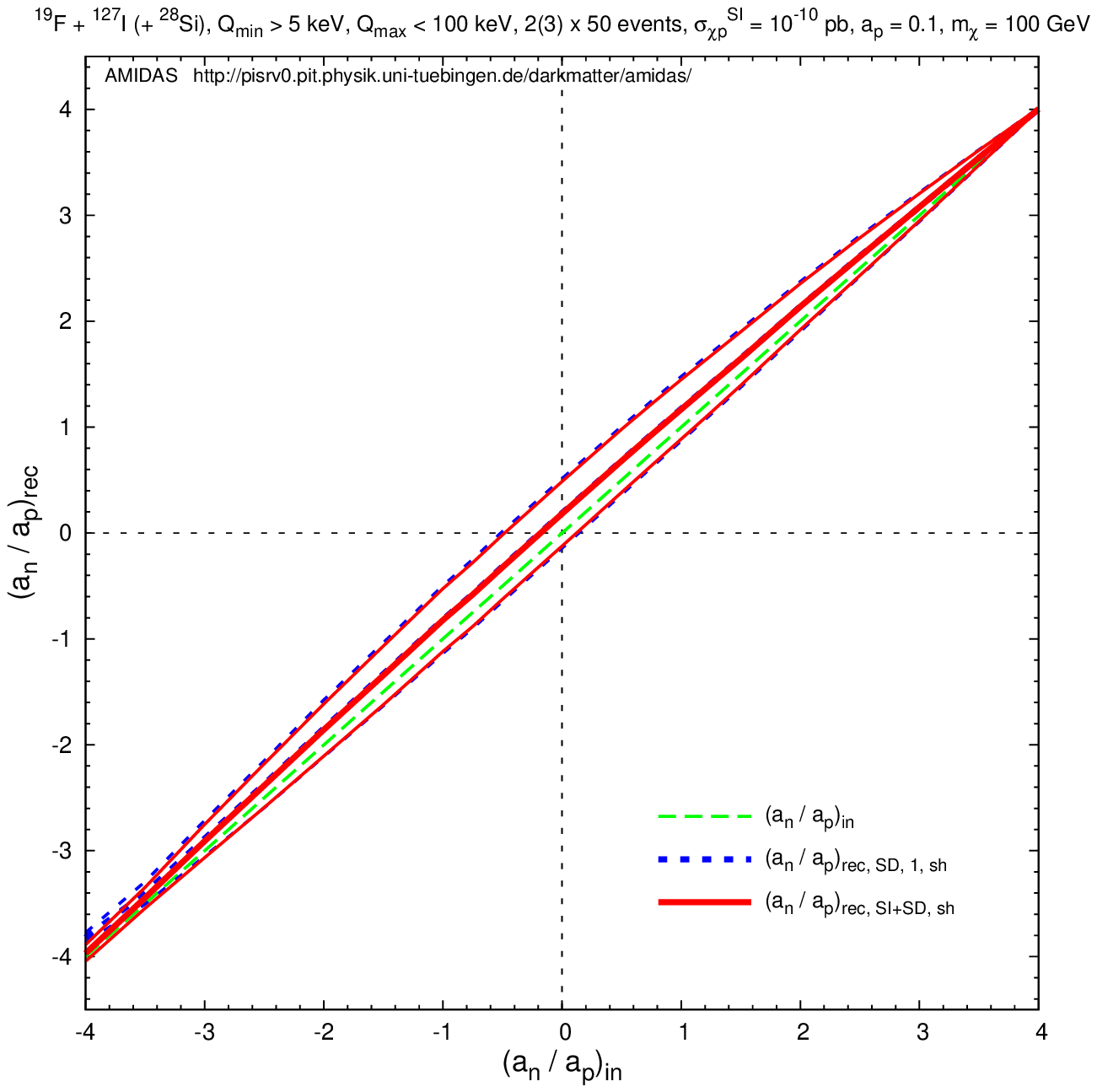}
\vspace{-0.75cm}
\end{center}
\caption{
 As in Figs.~\ref{fig:ranapSISD_ranap},
 except that $\rmXA{F}{19}$, $\rmXA{I}{127}$, and $\rmXA{Si}{28}$
 have been chosen as target nuclei.
 The scales used here are also different.
}
\label{fig:ranapSISD_ranap_fi}
\end{figure}
\begin{figure}[p!]
\begin{center}
\includegraphics[width=8.5cm]{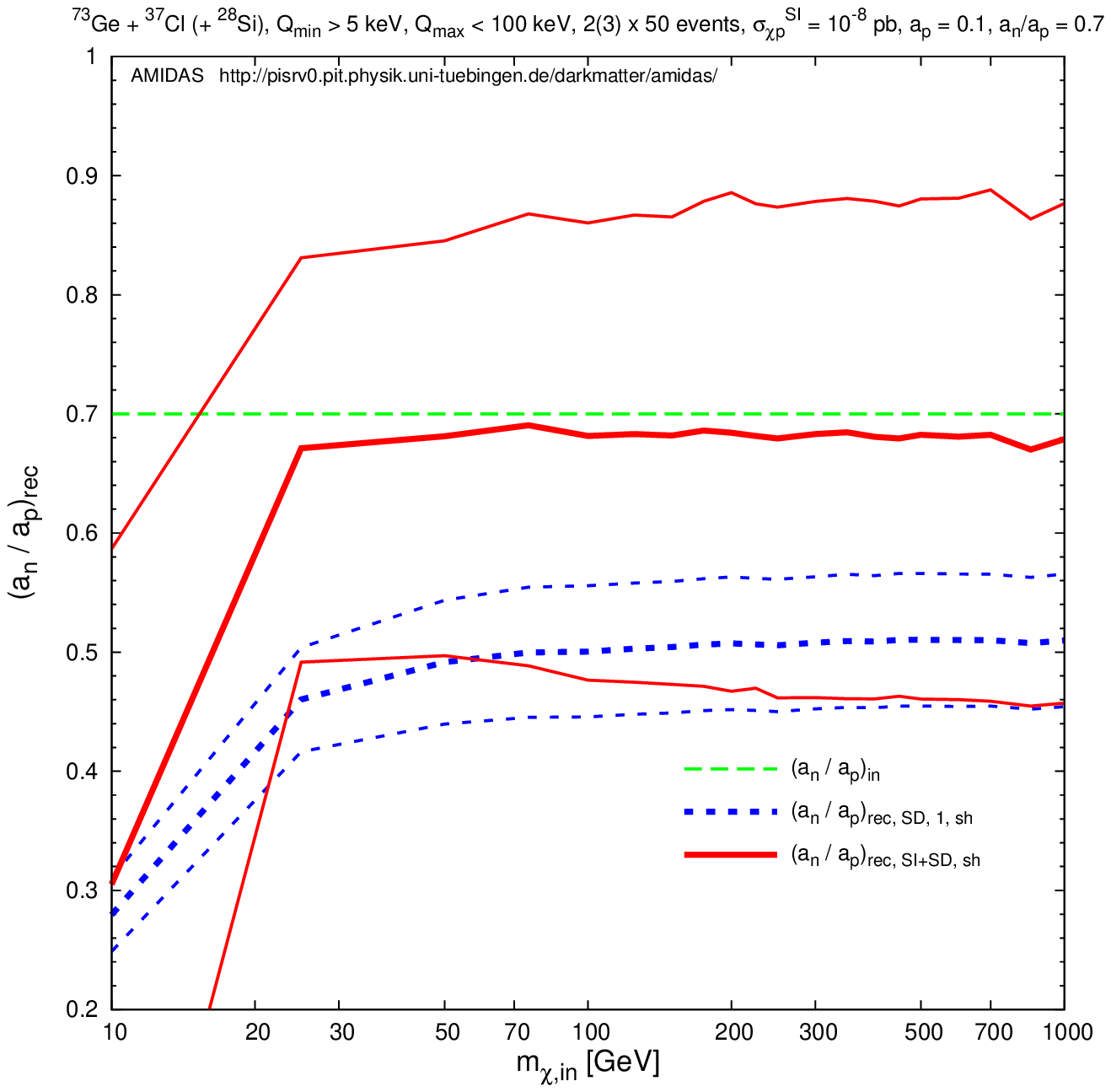}
\includegraphics[width=8.5cm]{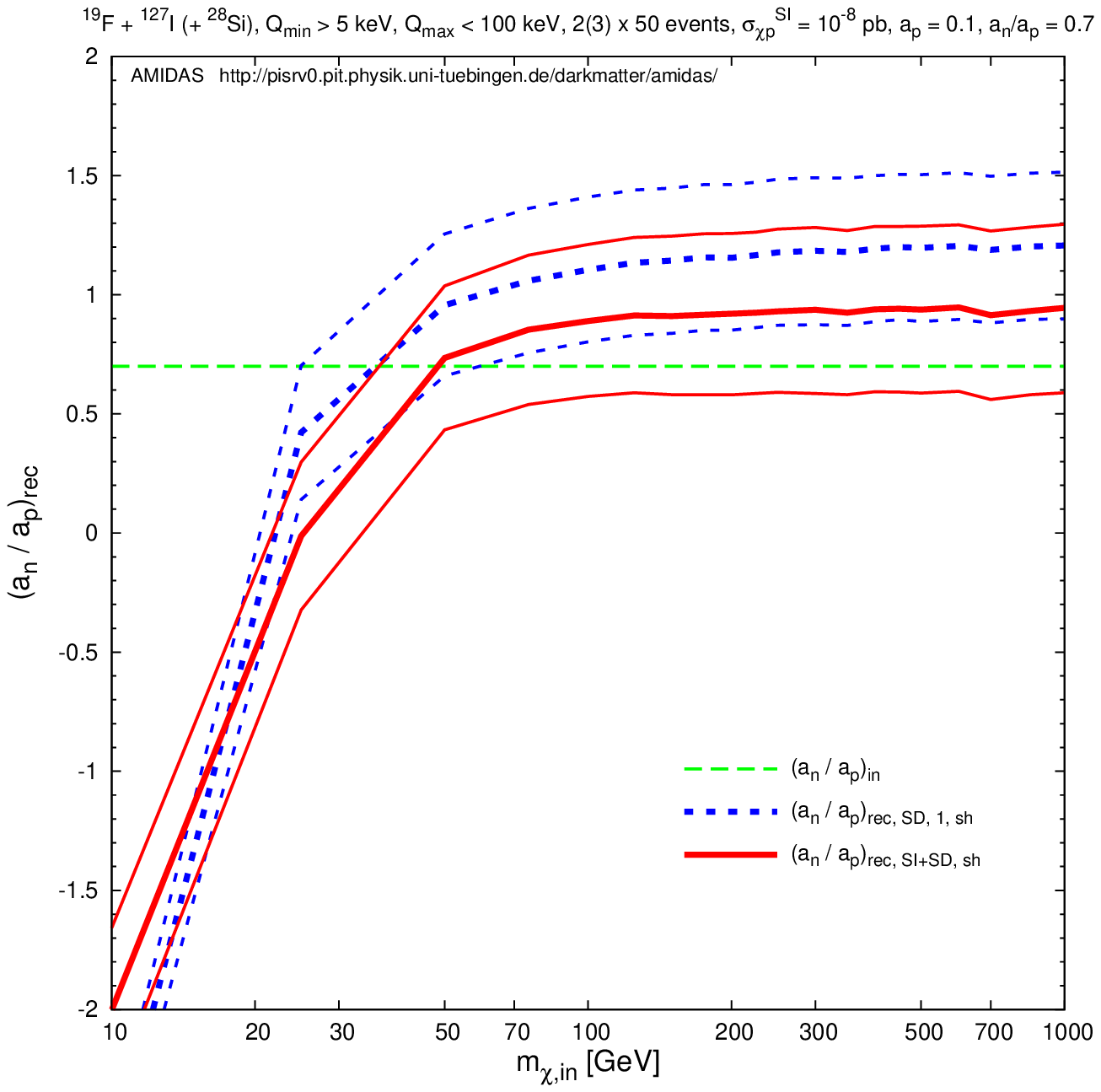}
\vspace{-0.75cm}
\end{center}
\caption{
 The reconstructed $\armn / \armp$ ratios
 estimated by Eqs.~(\ref{eqn:ranapSD}) (dashed blue)
 and (\ref{eqn:ranapSISD}) (solid red)
 and the lower and upper bounds of
 their 1$\sigma$ statistical uncertainties
 estimated by Eqs.~(\ref{eqn:sigma_ranapSD})
 and (\ref{eqn:sigma_ranapSISD})
 as functions of the input WIMP mass $\mchi$.
 The input $\armn / \armp$ ratio has been set as 0.7,
 the other parameters are as
 in Figs.~\ref{fig:ranapSISD_ranap} and \ref{fig:ranapSISD_ranap_fi}.
 Left:
 $\rmXA{Ge}{73}$, $\rmXA{Cl}{37}$, and $\rmXA{Si}{28}$
 have been chosen as the three target nuclei.
 Right:
 $\rmXA{F}{19}$, $\rmXA{I}{127}$, and $\rmXA{Si}{28}$
 have been chosen.
 Note here that
 the scales of the $(\armn / \armp)_{\rm rec}$--axes
 in two frames are different.
}
\label{fig:ranapSISD_mchi}
\end{figure}

 In Figs.~\ref{fig:ranapSISD_ranap_fi}
 the first two targets with both SI and SD sensitivities
 have been replaced again by $\rmXA{F}{19}$ and $\rmXA{I}{127}$.
 In contrast to Figs.~\ref{fig:ranapSISD_ranap},
 $\armn / \armp$ estimated by Eqs.~(\ref{eqn:ranapSD}) (dashed blue)
 and (\ref{eqn:ranapSISD}) (solid red)
 shown here are {\em overestimated},
 especially the ratio reconstructed
 under the assumption of a dominant SD interaction.
 Nevertheless,
 the $\armn / \armp$ ratio estimated by Eqs.~(\ref{eqn:ranapSISD})
 (solid red) in both Figs.~\ref{fig:ranapSISD_ranap}
 and \ref{fig:ranapSISD_ranap_fi} show that
 the ratio between two SD WIMP--nucleos couplings
 could in principle be estimated correctly
 with an $\sim 20 - 40\%$ statistical uncertainty
 {\em without} prior information on the WIMP mass
 {\em nor} on the SI WIMP--nucleon cross section.
 The (in)compatibility between
 the reconstructed $\armn / \armp$ ratios
 under different assumptions
 and/or with different combinations of target nuclei
 could also allow us to check
 whether the SD WIMP--nucleus interaction
 really dominates or not.

 Similar to the right frames of
 Figs.~\ref{fig:ranapSD_ranap_mchi}
 and \ref{fig:ranapSD_ranap_mchi_fi},
 Figs.~\ref{fig:ranapSISD_mchi} show
 the reconstructed $\armn / \armp$ ratios
 estimated by Eqs.~(\ref{eqn:ranapSD}) (dashed blue)
 and (\ref{eqn:ranapSISD}) (solid red)
 and the lower and upper bounds of
 their 1$\sigma$ statistical uncertainties
 estimated by Eqs.~(\ref{eqn:sigma_ranapSD})
 and (\ref{eqn:sigma_ranapSISD})
 as functions of the input WIMP mass $\mchi$.
 The over--/underestimated $\armn / \armp$ ratios
 with different combinations of target nuclei
 can be seen obviously here.
 For input WIMP masses $\mchi~\lsim~50$ GeV,
 all estimates are as usual (strongly) underestimated.
 Nevertheless,
 for WIMP masses $\mchi~\gsim~50$ GeV,
 the reconstructed 1$\sigma$ statistical uncertainty intervals
 estimated by Eqs.~(\ref{eqn:ranapSISD})
 and (\ref{eqn:sigma_ranapSISD}) (solid red)
 could basically cover the input (true) value
 pretty well.
\subsection{Choosing nuclei with \boldmath$\calCpY = 0$
            and $\SpX \gg \SnX \simeq 0$}
\begin{figure}[p!]
\begin{center}
\includegraphics[width=8.5cm]{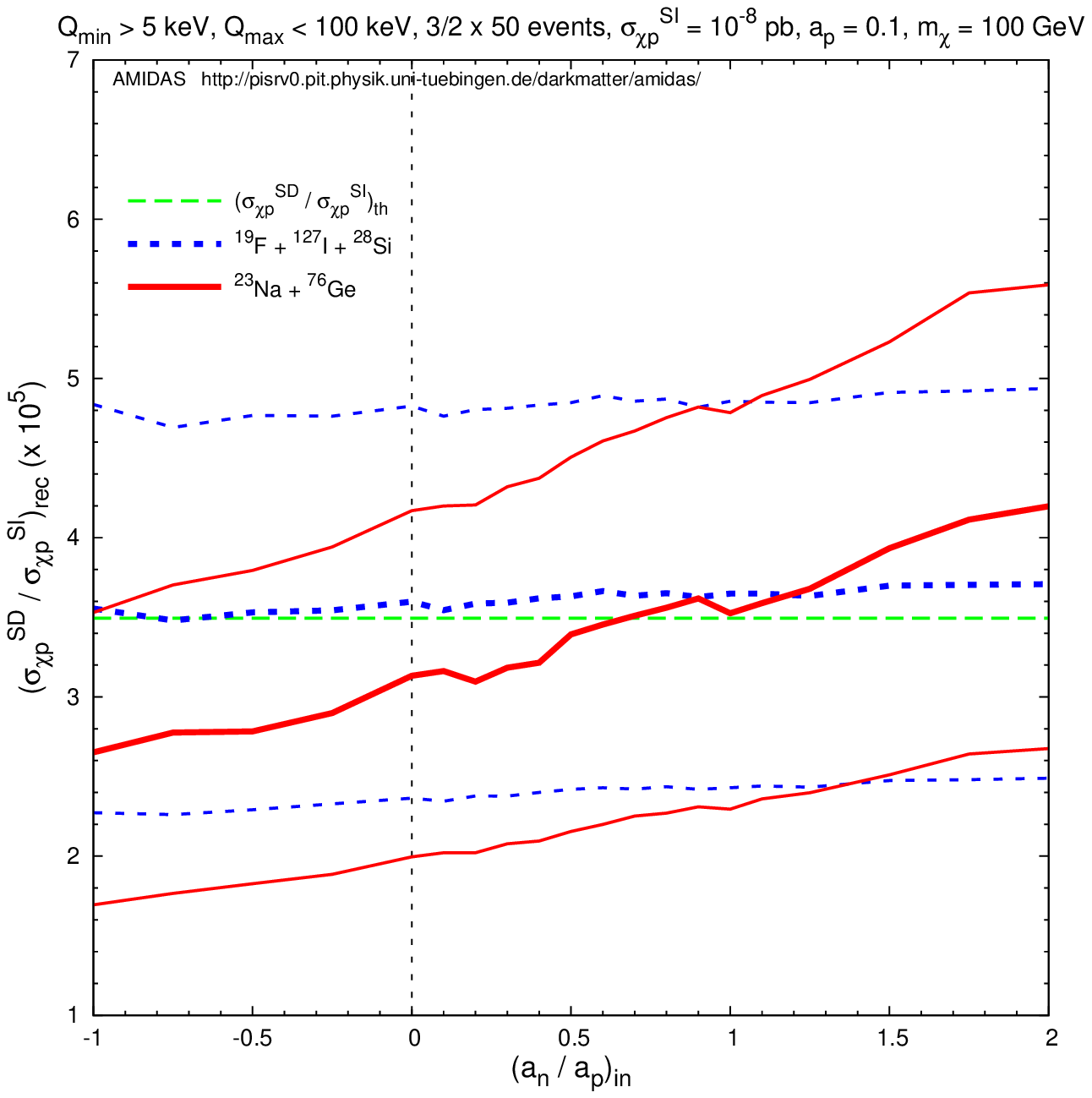}
\includegraphics[width=8.5cm]{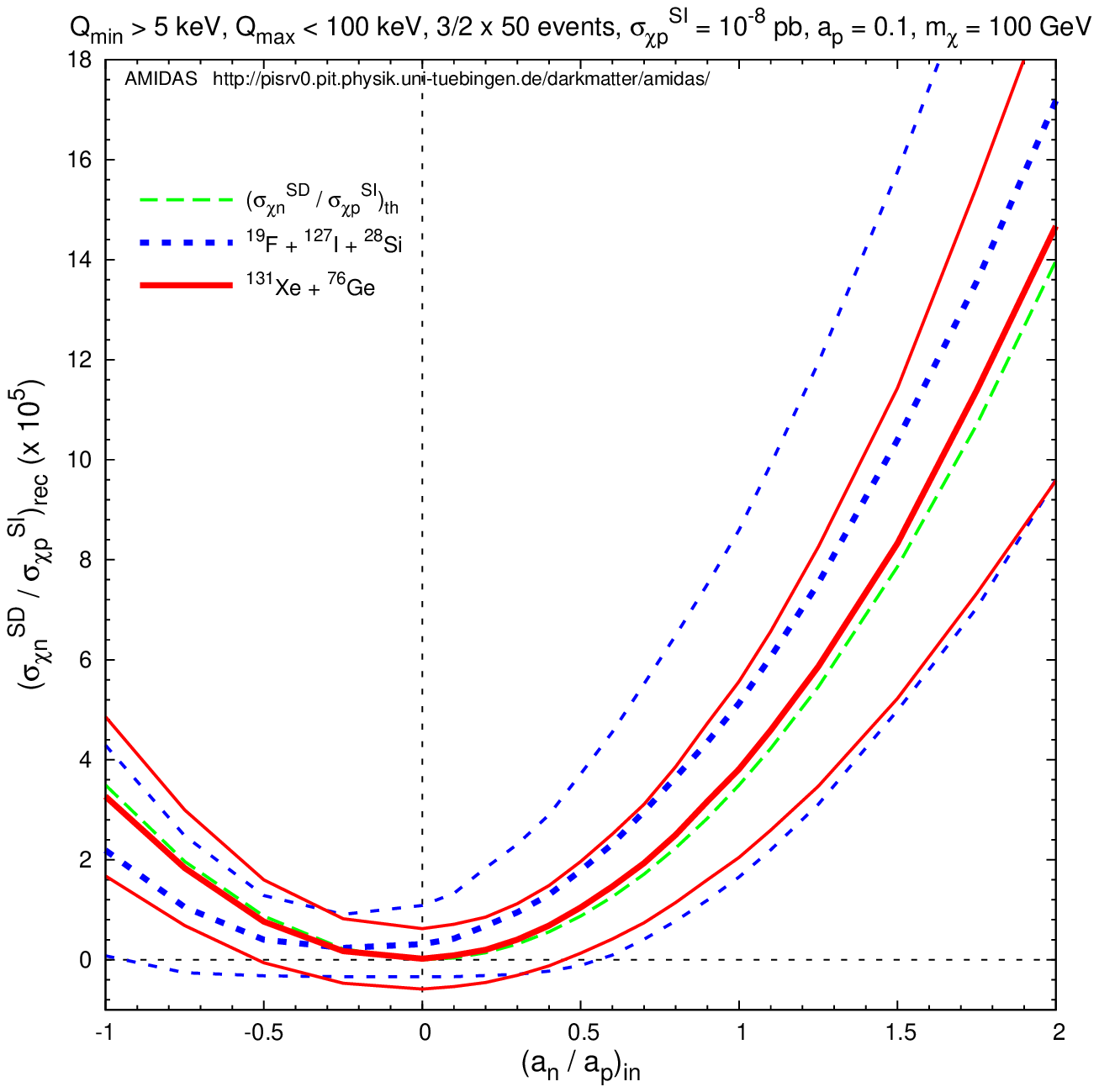}
\vspace{-1cm}
\end{center}
\caption{
 The reconstructed $\sigmapSD / \sigmapSI$ (left)
 and $\sigmanSD / \sigmapSI$ (right)
 and the lower and upper bounds of
 their 1$\sigma$ statistical uncertainties
 as functions of the input $\armn / \armp$,
 respectively.
 The dashed blue curves indicate the values
 estimated by Eq.~(\ref{eqn:rsigmaSDpSI})
 with $\armn / \armp$ estimated by Eq.~(\ref{eqn:ranapSISD})
 ({\em not} by Eq.~(\ref{eqn:ranapSD}));
 whereas the solid red curves indicate the values
 estimated by Eq.~(\ref{eqn:rsigmaSDpSI_even}).
 $\rmXA{Ge}{76}$ has been chosen as the second target
 with only an SI sensitivity and combined with
 $\rmXA{Na}{23}$ (for $\sigmapSD / \sigmapSI$) and
 $\rmXA{Xe}{131}$ (for $\sigmanSD / \sigmapSI$)
 for using Eq.~(\ref{eqn:rsigmaSDpSI_even}).
 Parameters are as in the left frame
 of Figs.~\ref{fig:ranapSISD_ranap_fi},
 the input SI WIMP--proton cross section
 has been set as $10^{-8}$ pb.
 Note that,
 since we fix $\sigmapSI$ and $\armp$,
 the theoretical curve of $\sigmapSD / \sigmapSI$
 is a constant,
 whereas the curve of $\sigmanSD / \sigmapSI \propto \armn^2$
 a parabola.
}
\label{fig:rsigmaSDSI_ranap_fi}
\end{figure}
\begin{figure}[p!]
\begin{center}
\includegraphics[width=8.5cm]{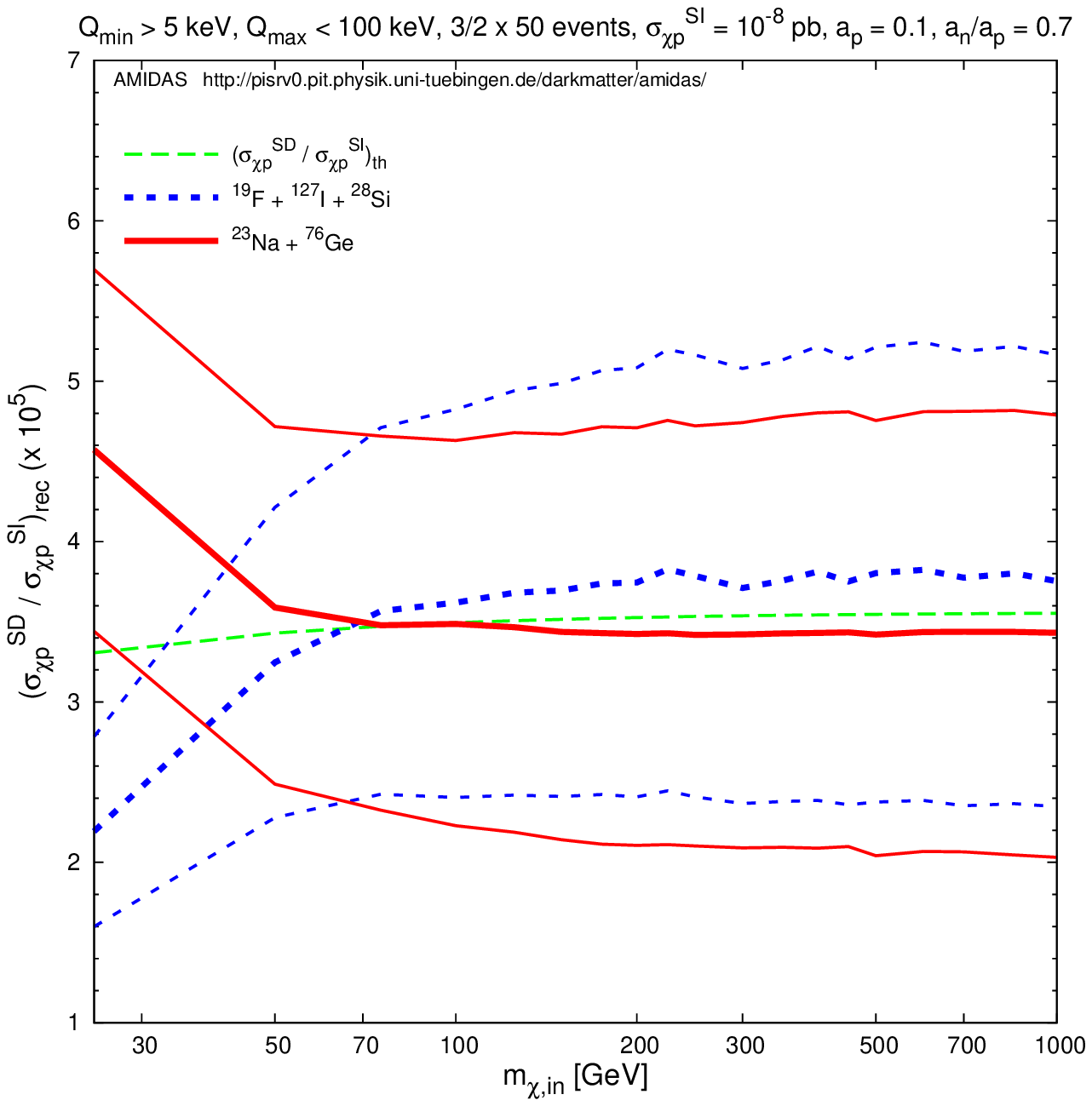}
\includegraphics[width=8.5cm]{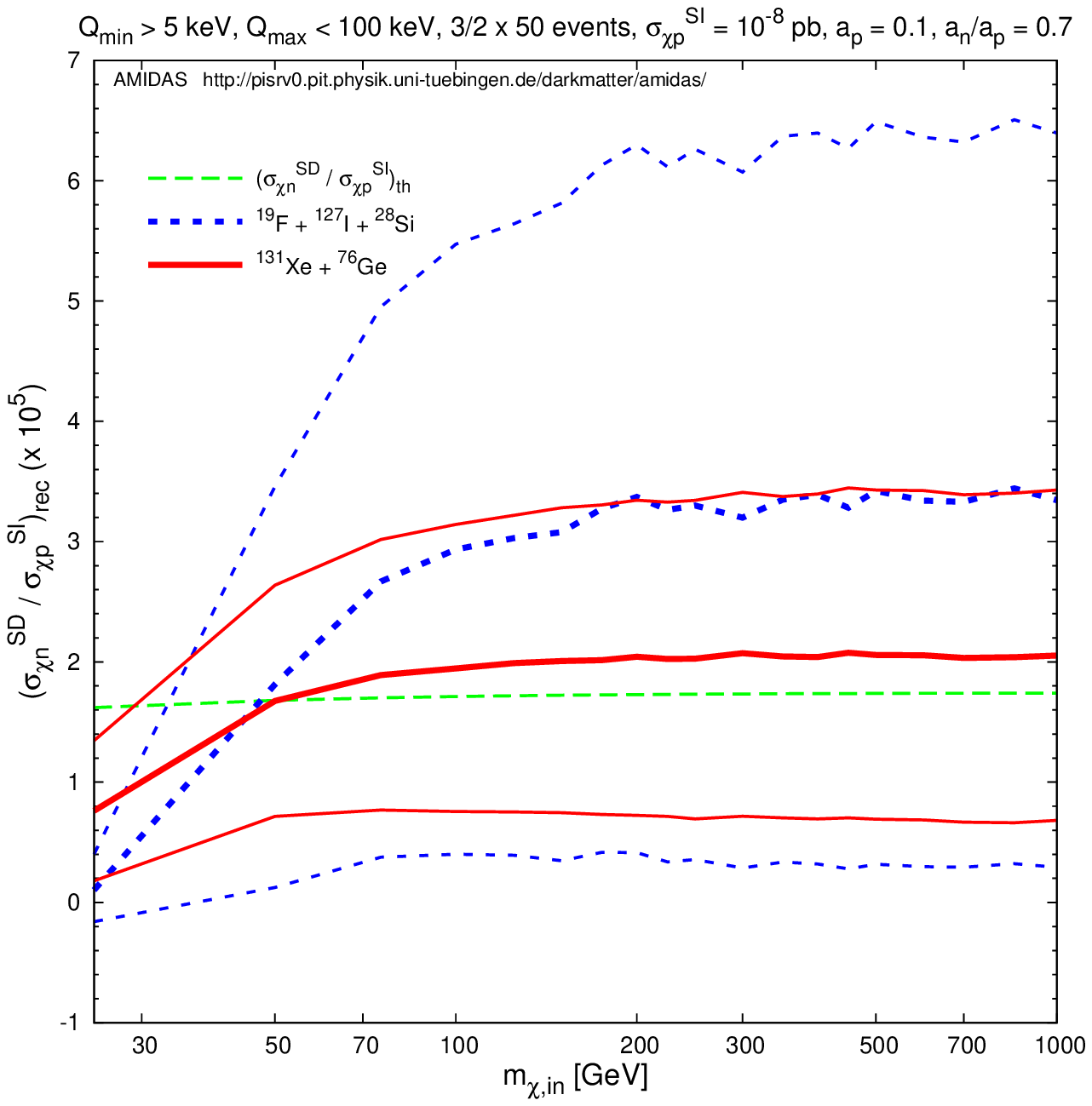}
\vspace{-1cm}
\end{center}
\caption{
 The reconstructed $\sigmapSD / \sigmapSI$ (left)
 and $\sigmanSD / \sigmapSI$ (right)
 and the lower and upper bounds of
 their 1$\sigma$ statistical uncertainties
 as functions of the input WIMP mass $\mchi$,
 respectively.
 The input SI WIMP--proton cross section
 has been set as $10^{-8}$ pb
 and the input $\armn / \armp = 0.7$,
 the other parameters and notations
 are as in Figs.~\ref{fig:rsigmaSDSI_ranap_fi}.
}
\label{fig:rsigmaSDSI_mchi_fi}
\end{figure}

 In the expression (\ref{eqn:rsigmaSDpSI})
 for the ratio of two WIMP--proton cross sections,
 there are four sources contributing statistical uncertainties,
 i.e., ${\cal C}_{{\rm p}, (X, Y)}$ and $\calR_{m, (X, Y)}$.
 In order to reduce the statistical uncertainty on
 the estimate of $\sigmapSD / \sigmapSI$,
 one can choose at first a nucleus
 with {\em only} an SI sensitivity
 as the second target:
\beq
   \SpY
 = \SnY
 = 0
\~,
\label{eqn:Sp/nY}
\eeq
 i.e.,
\beq
   {\cal C}_{{\rm p}, Y}
 = 0
\~.
\label{eqn:CpY}
\eeq
 The expression in Eq.~(\ref{eqn:rsigmaSDpSI})
 can thus be reduced to
 \cite{DMDDidentification-DMDE2009}
\beq
   \frac{\sigmapSD}{\sigmapSI}
 = \frac{\FSIQminY (\calR_{m, X} / \calR_{m, Y}) - \FSIQminX}
        {\calCpX \FSDQminX}
\~.
\label{eqn:rsigmaSDpSI_even}
\eeq
 Then we choose a nucleus with
 a (much) larger proton group spin
 as the first target:
\beq
        \SpX
 \gg    \SnX
 \simeq 0
\~,
\label{eqn:Sp/nX}
\eeq
 in order to eliminate the $\armn/\armp$ dependence of
 $\calCpX$ given in Eq.~(\ref{eqn:Cp})%
\footnote{
 One can also choose $\SnX \gg \SpX \simeq 0$
 and $\calCnX$ given in Eq.~(\ref{eqn:Cn}) becomes
\beq
        \calCnX
 \simeq \frac{4}{3} \afrac{J_X + 1}{J_X} \bfrac{\SnX}{A_X}^2
\~.
\label{eqn:CnX_n}
\eeq
}:
\beq
        \calCpX
 \simeq \frac{4}{3} \Afrac{J_X + 1}{J_X} \bfrac{\SpX}{A_X}^2
\~,
\label{eqn:CpX_p}
\eeq
 and the statistical uncertainty
 given in Eq.~(\ref{eqn:sigma_rsigmaSDpSI_ranapSISD})
 can be reduced to
\beq
        \sigma\afrac{\sigmapSD}{\sigmapSI}
 \simeq \frac{\FSIQminY (\calR_{m, X} / \calR_{m, Y})}{\calCpX \FSDQminX}
        \bbrac{  \frac{\sigma^2(r_X(\QminX))}{r_X^2(\QminX)}
               + \frac{\sigma^2(r_Y(\QminY))}{r_Y^2(\QminY)}}^{1/2}
\~.
\label{eqn:sigma_rsigmaSDpSI_ranapSISD_even}
\eeq

 In Figs.~\ref{fig:rsigmaSDSI_ranap_fi}
 I show the reconstructed $\sigmapSD / \sigmapSI$ (left)
 and $\sigmanSD / \sigmapSI$ (right)
 as functions of the input $\armn / \armp$,
 respectively.
 The dashed blue curves indicate the values
 estimated by Eq.~(\ref{eqn:rsigmaSDpSI})
 with $\armn / \armp$ estimated by Eq.~(\ref{eqn:ranapSISD})
 ({\em not} by Eq.~(\ref{eqn:ranapSD}));
 whereas the solid red curves indicate the values
 estimated by Eq.~(\ref{eqn:rsigmaSDpSI_even}).
 Since,
 as shown in the left frame of Figs.~\ref{fig:ranapSISD_ranap_fi},
 $\armn / \armp$ can be estimated pretty well
 by Eq.~(\ref{eqn:ranapSISD}) with the target combination F and I
 in the range of interest $-1 \le |\armn / \armp| \le 2$,
 $\sigmapSD / \sigmapSI$ shown in the left frame here
 can be reconstructed
 with an \mbox{$\sim$ 40\%} statistical uncertainty
 by using the combination of
 $\rmXA{F}{19}$ + $\rmXA{I}{127}$ + $\rmXA{Si}{28}$.
 On the other hand,
 the right frames of Figs.~\ref{fig:rsigmaSDSI_ranap_fi}
 and \ref{fig:rsigmaSDSI_mchi_fi}
 show also that
 $\sigmanSD / \sigmapSI$ could still be estimated well
 with $\rmXA{Xe}{131}$ and $\rmXA{Ge}{76}$
 by Eq.~(\ref{eqn:rsigmaSDpSI_even}),
 although the statistical uncertainty is now larger
 (\mbox{$\sim$ 70\%}).
\section{Estimating ratios of the SI WIMP--nucleon couplings}
 So far I have used the theoretical prediction
 (\ref{eqn:fp/n}) that
 the SI scalar WIMP coupling on protons
 is approximately equal to the coupling on neutrons.
 For the sake of completeness,
 I consider in this section briefly the case that
 WIMPs have different SI scalar or vector couplings
 on protons and on neutrons
 \cite{Feng:2011vu}.
 For WIMPs having only the scalar interaction with nuclei,
 the expression (\ref{eqn:sigma0_scalar})
 for $\sigmaSI$ can be rewritten as
\beq
   \sigmaSI
 = \afrac{4}{\pi} m_{\rm r, N}^2
   A^2 \bbrac{\afrac{Z}{A} f_{\rm p} + \abrac{1 - \frac{Z}{A}} f_{\rm n}}^2
\~.
\label{eqn:sigma0_scalar'}
\eeq
 Thus one can obtain the following replacements:
\beq
      \frac{J + 1}{J}
 \lto A^2
\~,
\label{eqn:J_A2}
\eeq
 and
\beq
      \Srmp
 \lto \frac{Z}{A}
\~,
      ~~~~~~~~~~~~~~~~ 
      \Srmn
 \lto 1 - \frac{Z}{A}
\~.
\eeq
 Substituting Eq.~(\ref{eqn:J_A2}) into Eq.~(\ref{eqn:RJnX}),
 we can get
\beq
   \calR_{J, n, X}^{\rm SI}
 = \frac{1}{A_X} \sfrac{\calR_{\sigma, X}}{\calR_{n, X}}
\~,
\label{eqn:RJnX_scalar}
\eeq
 where $\calR_{\sigma, X}$ and $\calR_{n, X}$
 are given in Eqs.~(\ref{eqn:RsigmaX_min}) and (\ref{eqn:RnX_min}).%
\footnote{
 Remind that
 the form factor $\FQ$ here
 must be chosen for the SI cross section.
}
 Then the ratio between the scalar WIMP coupling
 on protons and on neutrons
 can be estimated analogously to Eq.~(\ref{eqn:ranapSD}) as
\beq
   \afrac{f_{\rm n}}{f_{\rm p}}_{\pm, n}
 =-\frac{Z_X \pm Z_Y
         \sqrt{\calR_{\sigma, X} / \calR_{n, X}}
         \sqrt{\calR_{n, Y} / \calR_{\sigma, Y}}}
        {(A_X - Z_X) \pm (A_Y - Z_Y)
         \sqrt{\calR_{\sigma, X} / \calR_{n, X}}
         \sqrt{\calR_{n, Y} / \calR_{\sigma, Y}}}
\~,
\label{eqn:rfnfp}
\eeq
 with the following statistical uncertainty:
\beqn
        \sigma\abrac{\afrac{f_{\rm n}}{f_{\rm p}}_{\pm, n}}
 \=     \frac{\vBig{A_X Z_Y - A_Y Z_X} 
              \sqrt{\calR_{\sigma, X} / \calR_{n, X}}
              \sqrt{\calR_{n, Y} / \calR_{\sigma, Y}}}
             {2
              \bbigg{    (A_X - Z_X)
                    \pm (A_Y - Z_Y) 
                        \sqrt{\calR_{\sigma, X} / \calR_{n, X}}
                        \sqrt{\calR_{n, Y} / \calR_{\sigma, Y}} }^2}
        \non\\
 \conti ~~ \times
        \cBiggl{  \sum_{i, j = 1}^3
                  \bbrac{  \frac{1}{\calR_{n     , X}} \aPp{\calR_{n     , X}}{c_{i, X}}
                         - \frac{1}{\calR_{\sigma, X}} \aPp{\calR_{\sigma, X}}{c_{i, X}} } }
        \non\\
 \conti ~~~~~~~~~~~~~~~~ \times 
                  \bbrac{  \frac{1}{\calR_{n     , X}} \aPp{\calR_{n     , X}}{c_{j, X}}
                         - \frac{1}{\calR_{\sigma, X}} \aPp{\calR_{\sigma, X}}{c_{j, X}} }
                  {\rm cov}(c_{i, X}, c_{j, X})
        \non\\
 \conti ~~~~~~~~~~~~ 
        \cBiggr{+ (X \lto Y)}^{1/2}
\~.
\label{eqn:sigma_rfnfp}
\eeqn
 Note that,
 firstly,
 since $A - Z > 0$ for all nuclei,
 the inner solution of $f_{\rm n} / f_{\rm p}$
 given in Eq.~(\ref{eqn:rfnfp}) with
 a much smaller statistical uncertainty
 is always the ``$+$'' solution.
 Secondly,
 the two coincident points of the ``$+$'' and ``$-$'' soulutions
 decided by $-Z_X / (A_X - Z_X)$ and $-Z_Y / (A_Y - Z_Y)$
 are however always negative.
 While,
 for lighter nuclei, e.g.~$\rmXA{Si}{28}$ and $\rmXA{F}{19}$,
 the values of $- Z / (A - Z)$ are $\sim -1$;
 for heavier nuclei, e.g.~$\rmXA{I}{127}$ or $\rmXA{Xe}{131}$,
 these values are $\sim -0.7$.
 This means that,
 unfortunately,
 for confirming the $f_{\rm n} / f_{\rm p}$ ratio
 with the theoretical predicted value of $\sim 1$,
 we can only use
 the ``outer ($-$)'' solutions given in Eq.~(\ref{eqn:rfnfp})
 with much larger statistical uncertainties
 and data sets with piles of events
 should therefore be required.

 On the other hand,
 assuming that
 WIMPs have only the vector interaction with nuclei,
 according to the expression (\ref{eqn:sigma0_vector})
 for $\sigma_0^{\rm vector}$,
 we can write down the expression for
 the relative strength of two ``vector'' couplings directly as
\beq
   \afrac{b_{\rm n}}{b_{\rm p}}_{\pm, n}
 =-\frac{2 Z_X \pm 2 Z_Y
         \sqrt{\calR_{\sigma, X} / \calR_{n, X}}
         \sqrt{\calR_{n, Y} / \calR_{\sigma, Y}}}
        {(A_X - Z_X) \pm (A_Y - Z_Y)
         \sqrt{\calR_{\sigma, X} / \calR_{n, X}}
         \sqrt{\calR_{n, Y} / \calR_{\sigma, Y}}}
\~.
\label{eqn:rbnbp}
\eeq
 with the following statistical uncertainty:
\beqn
        \sigma\abrac{\afrac{b_{\rm n}}{b_{\rm p}}_{\pm, n}}
 \=     \frac{\vBig{A_X Z_Y - A_Y Z_X} 
              \sqrt{\calR_{\sigma, X} / \calR_{n, X}}
              \sqrt{\calR_{n, Y} / \calR_{\sigma, Y}}}
             {\bbrac{(A_X - Z_X) \pm (A_Y - Z_Y) 
              \sqrt{\calR_{\sigma, X} / \calR_{n, X}}
              \sqrt{\calR_{n, Y} / \calR_{\sigma, Y}}}^2}
        \non\\
 \conti ~~ \times
        \cBiggl{  \sum_{i, j = 1}^3
                  \bbrac{  \frac{1}{\calR_{n     , X}} \aPp{\calR_{n     , X}}{c_{i, X}}
                         - \frac{1}{\calR_{\sigma, X}} \aPp{\calR_{\sigma, X}}{c_{i, X}} } }
        \non\\
 \conti ~~~~~~~~~~~~~~~~ \times 
                  \bbrac{  \frac{1}{\calR_{n     , X}} \aPp{\calR_{n     , X}}{c_{j, X}}
                         - \frac{1}{\calR_{\sigma, X}} \aPp{\calR_{\sigma, X}}{c_{j, X}} }
                  {\rm cov}(c_{i, X}, c_{j, X})
        \non\\
 \conti ~~~~~~~~~~~~ 
        \cBiggr{+ (X \lto Y)}^{1/2}
\~.
\label{eqn:sigma_rbnbp}
\eeqn
 Note that
 the factor ``2'' appearing in the denominator of the prefacor
 in Eq.~(\ref{eqn:sigma_rfnfp})
 has been cancelled here. 
\section{Summary and conclusions}
 In this paper,
 I presented methods for determining
 ratios between different WIMP--nucleon couplings/cross sections
 from elastic WIMP--nucleus scattering experiments.
 All methods presented here
 are independent of the model of halo WIMPs
 as well as (practically) of the as yet unknown WIMP mass.
 Assuming that
 an exponential--like shape of the recoil spectrum
 is confirmed from experimental data,
 the required information
 are only the measured recoil energies
 and the number of events in the first energy bin
 from at least two direct detection experiments
 with different detector materials
 having spin sensitivities contributed
 from protons and/or from neutrons.
 Even better,
 our simulations show that,
 for estimating the relative strengths of
 different WIMP--nucleon couplings,
 one would only need events
 in the lowest available energy ranges.

 In order to avoid
 the uncertainty on the local WIMP density $\rho_0$,
 our analyses are based on combining two (or more) experiments
 using different target nuclei.
 By assuming,
 as the first step,
 that
 the SD WIMP--nucleus interaction dominates
 over the SI one,
 the expression for determining
 the ratio between two SD WIMP--nucleon couplings,
 $\armn / \armp$,
 has been rederived
 \cite{DMDDidentification-DMDE2009}.
 Then our simulations with different
 combinations of target nuclei show that,
 in order to obtain an unambiguous result
 with much smaller statistical uncertainty
 in the range of interest: $|\armn/\armp|~\lsim~2$,
 nuclei with sensitivities
 on {\em both} protons {\em and} neutrons
 should be more suitable than nuclei
 being sensitive (almost) only on protons or on neutrons.

 More generally,
 I considered also the combination
 of the SI and SD WIMP--nucleus cross sections.
 By using three different targets,
 two of them have non--zero group spins
 from protons and/or from neutrons,
 the second expression for determining
 the ratio between two SD WIMP--nucleon couplings
 can be rederived
 \cite{DMDDidentification-DMDE2009}.
 Although its statistical uncertainty
 depends on the relative strength between
 the SD and SI WIMP--nucleus interactions,
 the (in)compatibility between
 the $\armn / \armp$ ratio
 reconstructed under different assumptions
 and/or with different combinations of target nuclei
 could allow us to check whether
 the SD WIMP--nucleus interaction really dominates.
 Moreover,
 by using two or three different nuclei,
 one or two of them have non--zero group spins
 from protons and/or from neutrons,
 one can in principle also determine
 the ratios of the WIMP--proton/neutron cross sections
 to the SI ones,
 $\sigmapSD / \sigmapSI$ and $\sigmanSD / \sigmapSI$,
 directly.

 Our simulations presented here are based on
 several simplified assumptions.
 Firstly,
 the sample to be analyzed contains only signal events,
 i.e., is free of background%
\footnote{
 For background discrimination techniques and status
 in currently running and projected direct detection experiments
 see e.g.,
 \cite{Aprile09a, CRESST-bg, EDELWEISS-bg, Ahmed09b}.
}$^{,}$
\footnote{
 For detailed simulations and discussions about
 effects of residue background events
 on the determinations of ratios
 between different WIMP couplings/cross sections see
 \cite{DMDDbg-ranap}.
}.
 Secondly,
 all experimental systematic uncertainties as well as
 the uncertainty on the measurement of the recoil energy
 have been ignored.
 The energy resolution of most
 currently running and projected detectors
 is so good that its uncertainty can be neglected
 compared to the statistical uncertainty
 with (very) few events in the foreseeable future.

 In summary,
 I demonstrated in this paper the use of our new methods
 for extracting information on WIMP--nucleon couplings/cross sections,
 which are independent of models of WIMPs from particle physics
 as well as of models of the Galactic halo from cosmology.
 By combining with information on
 the estimation of the SI WIMP--nucleon coupling
 \cite{DMDDfp2-IDM2008, DMDDfp2},
 one could in principle estimate
 the absolute values of the spin--dependent couplings/cross sections.
 These information
 could help us not only to give constraints
 on different models of particle physics
 in the parameter space,
 but also to understand the nature of halo Dark Matter particles
 as well as to distinguish them
 between candidates predicted in different scenarios
 \cite{Bertone07, Barger08, Belanger08, Cotta09}.
\subsubsection*{Acknowledgments}
 The author appreciates M.~Drees and M.~Kakizaki
 for useful discussions
 and detailed comments on the preliminary draft.
 The author would also like to thank
 the Physikalisches Institut der Universit\"at T\"ubingen
 for the technical support of the computational work
 presented in this article.
 This work
 was partially supported by
 the National Science Council of R.O.C.~%
 under contract no.~NSC-99-2811-M-006-031
 and
 the LHC Physics Focus Group,
 National Center of Theoretical Sciences, R.O.C..

\appendix
\setcounter{equation}{0}
\setcounter{figure}{0}
\renewcommand{\theequation}{A\arabic{equation}}
\renewcommand{\thefigure}{A\arabic{figure}}
%
%
%
\section{Lists of needed formulae}
 Here I list all formulae needed
 for our model--independent data analyses
 described in this article.
 Detailed derivations and discussions
 can be found in Refs.~\cite{DMDDf1v, DMDDmchi}.
\subsection{Estimating \boldmath$r(\Qmin)$ and $I_n(\Qmin, \Qmax)$}
 First,
 consider experimental data described by
\beq
     {\T Q_n - \frac{b_n}{2}}
 \le \Qni
 \le {\T Q_n + \frac{b_n}{2}}
\~,
     ~~~~~~~~~~~~ 
     i
 =   1,~2,~\cdots,~N_n,~
     n
 =   1,~2,~\cdots,~B.
\label{eqn:Qni}
\eeq
 Here the total energy range between $\Qmin$ and $\Qmax$
 has been divided into $B$ bins
 with central points $Q_n$ and widths $b_n$.
 In each bin,
 $N_n$ events will be recorded.
 Since the recoil spectrum $dR / dQ$ is expected
 to be approximately exponential,
 the following ansatz for the {\em measured} recoil spectrum
 ({\em before} normalized by the experimental exposure $\calE$)
 in the $n$th bin has been introduced \cite{DMDDf1v}:
\beq
        \adRdQ_{{\rm expt}, \~ n}
 \equiv \adRdQ_{{\rm expt}, \~ Q \simeq Q_n}
 \equiv \rn  \~ e^{k_n (Q - Q_{s, n})}
\~.
\label{eqn:dRdQn}
\eeq
 Here $r_n$ is the standard estimator
 for $(dR / dQ)_{\rm expt}$ at $Q = Q_n$:
\beq
   r_n
 = \frac{N_n}{b_n}
\~,
\label{eqn:rn}
\eeq
 $k_n$ is the logarithmic slope of
 the recoil spectrum in the $n$th $Q-$bin,
 which can be computed numerically
 from the average value of the measured recoil energies
 in this bin:
\beq
   \bQn
 = \afrac{b_n}{2} \coth\afrac{k_n b_n}{2}-\frac{1}{k_n}
\~,
\label{eqn:bQn}
\eeq
 where
\beq
        \bQxn{\lambda}
 \equiv \frac{1}{N_n} \sumiNn \abrac{\Qni - Q_n}^{\lambda}
\~.
\label{eqn:bQn_lambda}
\eeq
 The error on the logarithmic slope $k_n$
 can be estimated from Eq.~(\ref{eqn:bQn}) directly as
\beq
   \sigma^2(k_n)
 = k_n^4
   \cbrac{  1
          - \bfrac{k_n b_n / 2}{\sinh (k_n b_n / 2)}^2}^{-2}
            \sigma^2\abrac{\bQn}
\~,
\label{eqn:sigma_kn}
\eeq
 with
\beq
   \sigma^2\abrac{\bQn}
 = \frac{1}{N_n - 1} \bbigg{\bQQn - \bQn^2}
\~.
\label{eqn:sigma_bQn}
\eeq
 $Q_{s, n}$ in the ansatz (\ref{eqn:dRdQn})
 is the shifted point at which
 the leading systematic error due to the ansatz
 is minimal \cite{DMDDf1v},
\beq
   Q_{s, n}
 = Q_n + \frac{1}{k_n} \ln\bfrac{\sinh(k_n b_n/2)}{k_n b_n/2}
\~.
\label{eqn:Qsn}
\eeq
 Note that $Q_{s, n}$ differs from
 the central point of the $n$th bin, $Q_n$.
 From the ansatz (\ref{eqn:dRdQn}),
 the counting rate at $Q = \Qmin$ can be calculated by
\beq
   r(\Qmin)
 = r_1 e^{k_1 (\Qmin - Q_{s, 1})}
\~,
\label{eqn:rmin_eq}
\eeq
 and its statistical error can be expressed as
\beq
   \sigma^2(r(\Qmin))
 = r^2(\Qmin) 
   \cbrac{  \frac{1}{N_1}
          + \bbrac{  \frac{1}{k_1}
                   - \afrac{b_1}{2} 
                     \abrac{  1
                            + \coth\afrac{b_1 k_1}{2}}}^2
            \sigma^2(k_1)}
\~,
\label{eqn:sigma_rmin}
\eeq
 since
\beq
   \sigma^2(r_n)
 = \frac{N_n}{b_n^2}
\~.
\label{eqn:sigma_rn}
\eeq
 Finally,
 since all $I_n$ are determined from the same data,
 they are correlated with
\beq
   {\rm cov}(I_n, I_m)
 = \sum_{a = 1}^{N_{\rm tot}} \frac{Q_a^{(n+m-2)/2}}{F^4(Q_a)}
\~,
\label{eqn:cov_In}
\eeq
 where the sum runs over all events
 with recoil energy between $\Qmin$ and $\Qmax$. 
 And the correlation between the errors on $r(\Qmin)$,
 which is calculated entirely
 from the events in the first bin,
 and on $I_n$ is given by
\beqn
 \conti {\rm cov}(r(\Qmin), I_n)
        \non\\
 \=     r(\Qmin) \~ I_n(\Qmin, \Qmin + b_1)
        \non\\
 \conti ~~~~ \times 
        \cleft{  \frac{1}{N_1} 
               + \bbrac{  \frac{1}{k_1}
                        - \afrac{b_1}{2} \abrac{1 + \coth\afrac{b_1 k_1}{2}}}}
        \non\\
 \conti ~~~~~~~~~~~~~~ \times 
        \cright{ \bbrac{  \frac{I_{n+2}(\Qmin, \Qmin + b_1)}
                               {I_{n  }(\Qmin, \Qmin + b_1)}
                        - Q_1
                        + \frac{1}{k_1}
                        - \afrac{b_1}{2} \coth\afrac{b_1 k_1} {2}}
        \sigma^2(k_1)}
\~;
\label{eqn:cov_rmin_In}
\eeqn
 note that
 the sums $I_i$ here only count in the first bin,
 which ends at $Q = \Qmin + b_1$.

 On the other hand,
 with a functional form of the recoil spectrum
 (e.g., fitted to experimental data),
 $(dR / dQ)_{\rm expt}$,
 one can use the following integral forms
 to replace the summations given above.
 Firstly,
 the average $Q-$value in the $n$th bin
 defined in Eq.~(\ref{eqn:bQn_lambda})
 can be calculated by
\beq
   \bQxn{\lambda}
 = \frac{1}{N_n} \intQnbn \abrac{Q - Q_n}^{\lambda} \adRdQ_{\rm expt} dQ
\~.
\label{eqn:bQn_lambda_int}
\eeq
 For $I_n(\Qmin, \Qmax)$ given in Eq.~(\ref{eqn:In_sum}),
 we have
\beq
   I_n(\Qmin, \Qmax)
 = \int_{\Qmin}^{\Qmax} \frac{Q^{(n-1)/2}}{F^2(Q)} \adRdQ_{\rm expt} dQ
\~,
\label{eqn:In_int}
\eeq 
 and similarly for the covariance matrix for $I_n$
 in Eq.~(\ref{eqn:cov_In}),
\beq
   {\rm cov}(I_n, I_m)
 = \int_{\Qmin}^{\Qmax} \frac{Q^{(n+m-2)/2}}{F^4(Q)} \adRdQ_{\rm expt} dQ
\~.
\label{eqn:cov_In_int}
\eeq 
 Remind that
 $(dR / dQ)_{\rm expt}$ is the {\em measured} recoil spectrum
 {\em before} normalized by the exposure.
 Finally,
 $I_i(\Qmin, \Qmin + b_1)$ needed in Eq.~(\ref{eqn:cov_rmin_In})
 can be calculated by
\beq
   I_n(\Qmin, \Qmin + b_1)
 = \int_{\Qmin}^{\Qmin + b_1}
   \frac{Q^{(n-1)/2}}{F^2(Q)} \bbigg{r_1 \~ e^{k_1 (Q - Q_{s, 1})}} dQ
\~.
\label{eqn:In_1_int}
\eeq 
 Note that,
 firstly,
 $r(\Qmin)$ and $I_n(\Qmin, \Qmin + b_1)$ should be
 estimated by Eqs.~(\ref{eqn:rmin_eq}) and (\ref{eqn:In_1_int})
 with $r_1$, $k_1$ and $Q_{s, 1}$
 estimated by Eqs.~(\ref{eqn:rn}), (\ref{eqn:bQn}), and (\ref{eqn:Qsn})
 in order to use the other formulae for estimating
 the (correlations between the) statistical errors
 without any modification.
 Secondly,
 $r(\Qmin)$ and $I_n(\Qmin, \Qmax)$ estimated
 from a scattering spectrum fitted to experimental data
 are usually not model--independent any more.
 Moreover,
 for the use of Eqs.~(\ref{eqn:In_sum}),
 (\ref{eqn:cov_In}), (\ref{eqn:In_int}),
 (\ref{eqn:cov_In_int}), and (\ref{eqn:In_1_int})
 the elastic nuclear form factor $\FQ$
 should be understood to be chosen
 for the SI and SD WIMP--nucleon cross section correspondingly.
\subsection{Derivatives of \boldmath$\calR_{n, X}$ and $\calR_{\sigma, X}$}
 First,
 from Eq.~(\ref{eqn:RnX_min}) one can find
 explicit expressions for the derivatives of $\calR_{n, X}$
 with respect to $c_{i, X}$ are:
\cheqnXa{A}
\beq
   \Pp{\calR_{n, X}}{\InX}
 = \frac{n + 1}{n}
   \bfrac{\FQminX}{2 \QminX^{(n + 1) / 2} r_X(\QminX) + (n + 1) \InX \FQminX}
   \calR_{n, X}
\~,
\label{eqn:dRnX_dInX}
\eeq
\cheqnXb{A}
\beq
   \Pp{\calR_{n, X}}{\IzX}
 =-\frac{1}{n}
   \bfrac{\FQminX}{2 \QminX^{1 / 2} r_X(\QminX) + \IzX \FQminX}
   \calR_{n, X}
\~,
\label{eqn:dRnX_dIzX}
\eeq
 and
\cheqnXc{A}
\beqn
        \Pp{\calR_{n, X}}{r_X(\QminX)}
 \=     \frac{2}{n}
        \bfrac{  \QminX^{(n + 1) / 2} \IzX        - (n + 1) \QminX^{1 / 2} \InX}
              {2 \QminX^{(n + 1) / 2} r_X(\QminX) + (n + 1) \InX \FQminX}
        \non\\
 \conti ~~~~~~~~~~~~~~~~ \times 
        \bfrac{\FQminX}{2 \QminX^{1 / 2} r_X(\QminX) + \IzX \FQminX}
        \calR_{n, X}
\~;
\label{eqn:dRnX_drminX}
\eeqn
\cheqnX{A}%
 explicit expressions for the derivatives of $\calR_{n, Y}$
 with respect to $c_{i, Y}$ can be given analogously.
 Note that,
 firstly,
 factors $\calR_{n, (X, Y)}$ appear in all these expressions,
 which can practically be cancelled by the prefactors
 in the bracket in Eq.~(\ref{eqn:sigma_ranapSD}).
 Secondly,
 all the $I_{0, (X, Y)}$ and $I_{n, (X, Y)}$ should be understood
 to be computed according to
 Eq.~(\ref{eqn:In_sum}) or (\ref{eqn:In_int})
 with integration limits $\Qmin$ and $\Qmax$
 specific for that target.

 Similarly,
 expressions for the derivatives of $\calR_{\sigma, X}$
 can be computed from Eq.~(\ref{eqn:RsigmaX_min}) as
\cheqnXa{A}
\beq
   \Pp{\calR_{\sigma, X}}{\IzX}
 = \bfrac{\FQminX}{2 \QminX^{1 / 2} r_X(\QminX) + \IzX \FQminX}
   \calR_{\sigma, X}
\~,
\label{eqn:dRsigmaX_dIzX}
\eeq
\cheqnXb{A}
\beq
   \Pp{\calR_{\sigma, X}}{r_X(\QminX)}
 = \bfrac{2 \QminX^{1 / 2}}{2 \QminX^{1 / 2} r_X(\QminX) + \IzX \FQminX}
   \calR_{\sigma, X}
\~;
\label{eqn:dRsigmaX_drminX}
\eeq
\cheqnX{A}%
 and similarly for the derivatives of $\calR_{\sigma, Y}$.
 Remind that
 factors $\calR_{\sigma, (X, Y)}$ appearing here
 can also be cancelled by the prefactors
 in the bracket in Eq.~(\ref{eqn:sigma_ranapSD}).
\subsection{Derivatives of \boldmath$\sigmapSD / \sigmapSI$}
 From the expression (\ref{eqn:rsigmaSDpSI})
 for estimating $\sigmapSD / \sigmapSI$,
 its derivatives with respect to ${\cal C}_{{\rm p}, (X, Y)}$
 can be given as
\cheqnXa{A}
\beq
   \pp{\calCpX} \afrac{\sigmapSD}{\sigmapSI}
 =-\frac{\FSDQminX \calR_{m, Y}}
        {\calCpX \FSDQminX \calR_{m, Y} - \calCpY  \FSDQminY \calR_{m, X}}
   \afrac{\sigmapSD}{\sigmapSI}
\~,
\label{eqn:drsigmaSDpSI_dCpX}
\eeq
 and
\cheqnXb{A}
\beq
   \pp{\calCpY} \afrac{\sigmapSD}{\sigmapSI}
 = \frac{\FSDQminY \calR_{m, X}}
        {\calCpX \FSDQminX \calR_{m, Y} - \calCpY  \FSDQminY \calR_{m, X}}
   \afrac{\sigmapSD}{\sigmapSI}
\~.
\label{eqn:drsigmaSDpSI_dCpY}
\eeq
\cheqnX{A}
 Meanwhile,
 the derivatives of $\sigmapSD / \sigmapSI$
 with respect to $\calR_{m, (X, Y)}$ are
\cheqnXa{A}
\beqn
 \conti \pp{\calR_{m, X}} \afrac{\sigmapSD}{\sigmapSI}
        \non\\
 \=    -\frac{\bBig{  \calCpX \FSDQminX \FSIQminY
                    - \calCpY \FSDQminY \FSIQminX}
              \calR_{m, Y} }
             {\bBig{  \calCpX \FSDQminX \calR_{m, Y}
                    - \calCpY \FSDQminY \calR_{m, X} }^2}
\~,
\label{eqn:drsigmaSDpSI_dRmX}
\eeqn
 and
\cheqnXb{A}
\beqn
 \conti \pp{\calR_{m, Y}} \afrac{\sigmapSD}{\sigmapSI}
        \non\\
 \=     \frac{\bBig{  \calCpY \FSDQminY \FSIQminX
                    - \calCpX \FSDQminX \FSIQminY}
              \calR_{m, X} }
             {\bBig{  \calCpX \FSDQminX \calR_{m, Y}
                    - \calCpY \FSDQminY \calR_{m, X} }^2}
\~.
\label{eqn:drsigmaSDpSI_dRmY}
\eeqn
\cheqnX{A}%
 On the other hand,
 from expression (\ref{eqn:Cp}) for $\calCp$
 one can find that
\beq
   \Pp{\calCp}{(\armn / \armp)}
 = \frac{2 \calCp}{\Srmp / \Srmn + \armn / \armp}
\~,
\label{eqn:dCp_dranap}
\eeq
 and,
 since we estimate in fact always $\armn / \armp$,
 one needs practically
\beq
   \Pp{\calCn}{(\armn / \armp)}
 =-\frac{2 \calCn}{\armn / \armp + (\Srmn / \Srmp) (\armn / \armp)^2}
\~.
\label{eqn:dCp_dranap}
\eeq
\subsection{Derivatives of \boldmath$(\armn / \armp)_{\pm}^{\rm SI + SD}$}
 At first,
 from the first and second lines of
 the expression (\ref{eqn:ranapSISD}),
 we have,
\cheqnXa{A}
\beqn
        \pp{\cpX}\afrac{\armn}{\armp}_{\pm}^{\rm SI + SD}
 \=    -\frac{1}{\abrac{\cpX \snpX^2 - \cpY \snpY^2}^2}
        \non\\
 \conti ~~~~~~ \times 
        \bBiggl{    \cpY \snpX \snpY \abrac{\snpX - \snpY}}
        \non\\
 \conti ~~~~~~~~~~~~~~~~ 
        \bBiggr{\pm \frac{1}{2} \sfrac{\cpY}{\cpX}
                    \abrac{\cpX \snpX^2 + \cpY \snpY^2}
                    \vbrac{\snpX - \snpY}}
        \non\\
 \= \cleft{\renewcommand{\arraystretch}{0.5}
           \begin{array}{l l l}
            \\
            \D \mp \frac{\sqrt{\cpX \cpY} \abrac{\snpX - \snpY}}
                        {2 \cpX \abrac{\sqrt{\cpX} \snpX \mp \sqrt{\cpY} \snpY}^2}\~, &
            ~~ & ({\rm for}~\snpX > \snpY), \\~\\~\\
            \D \pm \frac{\sqrt{\cpX \cpY} \abrac{\snpX - \snpY}}
                        {2 \cpX \abrac{\sqrt{\cpX} \snpX \pm \sqrt{\cpY} \snpY}^2}\~, &
               & ({\rm for}~\snpX < \snpY), \\~\\
           \end{array}}
\label{eqn:dranapSISD_dcpX}
\eeqn
 and
\cheqnXb{A}
\beqn
        \pp{\cpY}\afrac{\armn}{\armp}_{\pm}^{\rm SI + SD}
 \=     \frac{1}{\abrac{\cpX \snpX^2 - \cpY \snpY^2}^2}
        \non\\
 \conti ~~~~~~ \times 
        \bBiggl{    \cpX \snpX \snpY \abrac{\snpX - \snpY}}
        \non\\
 \conti ~~~~~~~~~~~~~~~~ 
        \bBiggr{\pm \frac{1}{2} \sfrac{\cpX}{\cpY}
                    \abrac{\cpX \snpX^2 + \cpY \snpY^2}
                    \vbrac{\snpX - \snpY}}
        \non\\
 \= \cleft{\renewcommand{\arraystretch}{0.5}
           \begin{array}{l l l}
            \\
            \D \pm \frac{\sqrt{\cpX \cpY} \abrac{\snpX - \snpY}}
                        {2 \cpY \abrac{\sqrt{\cpX} \snpX \mp \sqrt{\cpY} \snpY}^2}\~, &
            ~~ & ({\rm for}~\snpX > \snpY), \\~\\~\\
            \D \mp \frac{\sqrt{\cpX \cpY} \abrac{\snpX - \snpY}}
                        {2 \cpY \abrac{\sqrt{\cpX} \snpX \pm \sqrt{\cpY} \snpY}^2}\~, &
               & ({\rm for}~\snpX < \snpY). \\~\\
           \end{array}}
\label{eqn:dranapSISD_dcpY}
\eeqn
\cheqnX{A}
 Then,
 from the definition (\ref{eqn:cpX}) of $\cpX$,
 one can get directly
\cheqnXa{A}
\beq
   \Pp{\cpX}{r_X(\QminX)}
 = 0
\~,
\label{eqn:dcpX_drminX}
\eeq
\cheqnXb{A}
\beq
   \Pp{\cpX}{r_Y(\QminY)}
 = \FSIQminZ \FSDQminX \cdot \frac{4}{3} \Afrac{J_X + 1}{J_X} \bfrac{\SpX}{A_X}^2
   \cdot \frac{\calR_{m, YZ}}{r_Y(\QminY)}
\~,
\label{eqn:dcpX_drminY}
\eeq
 and
\cheqnXc{A}
\beq
   \Pp{\cpX}{r_Z(\QminZ)}
 =-\FSIQminZ \FSDQminX \cdot \frac{4}{3} \Afrac{J_X + 1}{J_X} \bfrac{\SpX}{A_X}^2
   \cdot \frac{\calR_{m, YZ}}{r_Z(\QminZ)}
\~.
\label{eqn:dcpX_drminZ}
\eeq
\cheqnX{A}
 Similarly, 
 from the definition (\ref{eqn:cpY}) of $\cpY$,
 we have
\cheqnXa{A}
\beq
   \Pp{\cpY}{r_X(\QminX)}
 = \FSIQminZ \FSDQminY \cdot \frac{4}{3} \Afrac{J_Y + 1}{J_Y} \bfrac{\SpY}{A_Y}^2
   \cdot \frac{\calR_{m, XZ}}{r_X(\QminX)}
\~,
\label{eqn:dcpY_drminX}
\eeq
\cheqnXb{A}
\beq
   \Pp{\cpY}{r_Y(\QminY)}
 = 0
\~,
\label{eqn:dcpY_drminY}
\eeq
 and
\cheqnXc{A}
\beq
   \Pp{\cpY}{r_Z(\QminZ)}
 =-\FSIQminZ \FSDQminY \cdot \frac{4}{3} \Afrac{J_Y + 1}{J_Y} \bfrac{\SpY}{A_Y}^2
   \cdot \frac{\calR_{m, XZ}}{r_Z(\QminZ)}
\~.
\label{eqn:dcpY_drminZ}
\eeq
\cheqnX{A}
\end{document}